\documentclass[iop]{emulateapj}
\usepackage{leftidx}
\usepackage{amsmath,amssymb,amsthm}
\usepackage{natbib,twoopt}
\usepackage{graphicx}
\usepackage{svg}
\usepackage{txfonts}
\usepackage{atbegshi}
\usepackage{hyperref}
\newcommand{\orcid}[1]{\unskip\protect\href{https://orcid.org/#1}{\protect\includegraphics[width=8pt,clip]{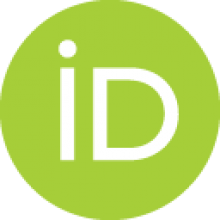}}}
\def\hackaltaffiltext#1#2{\AtBeginShipoutNext{\footnotetext[#1]{#2}\stepcounter{footnote}}}

\begin{document}

\newcommand{\kms}{km~s$^{-1}$}
\newcommand{\ms}{m~s$^{-1}$}
\newcommand{\teff}{$T_{\rm eff}$}
\newcommand{\logg}{log~$g$}
\newcommand{\mh}{[M/H]}
\newcommand{\feh}{[Fe/H]}
\newcommand{\am}{[$\alpha$/M]}
\newcommand{\cm}{[C/M]}
\newcommand{\ch}{[C/H]}
\newcommand{\cih}{[CI/H]}
\newcommand{\nm}{[N/M]}
\newcommand{\nh}{[N/H]}
\newcommand{\oh}{[O/H]}
\newcommand{\nah}{[Na/H]}
\newcommand{\mgh}{[Mg/H]}  
\newcommand{\alh}{[Al/H]} 
\newcommand{\sih}{[Si/H]} 
\newcommand{\ph}{[P/H]} 
\newcommand{\sh}{[S/H]}  
\newcommand{\kh}{[K/H]}  
\newcommand{\cah}{[Ca/H]}  
\newcommand{\tih}{[Ti/H]} 
\newcommand{\tiih}{[TiII/H]} 
\newcommand{\vh}{[V/H]} 
\newcommand{\crh}{[Cr/H]}  
\newcommand{\mnh}{[Mn/H]}  
\newcommand{\coh}{[Co/H]}  
\newcommand{\nih}{[Ni/H]}  
\newcommand{\cuh}{[Cu/H]}  
\newcommand{\ceh}{[Ce/H]}  
\newcommand{\ndh}{[Nd/H]}  
\newcommand{\xh}{[X/H]}  
\newcommand{\xm}{[X/M]}  

\newcommand{\om}{[O/M]}
\newcommand{\nam}{[Na/M]}
\newcommand{\mgm}{[Mg/M]}  
\newcommand{\alm}{[Al/M]} 
\newcommand{\simm}{[Si/M]} 
\newcommand{\pmm}{[P/M]} 
\newcommand{\sm}{[S/M]}  
\newcommand{\km}{[K/M]}  
\newcommand{\cam}{[Ca/M]}  
\newcommand{\tim}{[Ti/M]} 
\newcommand{\vm}{[V/M]} 
\newcommand{\crm}{[Cr/M]}  
\newcommand{\mnm}{[Mn/M]}  
\newcommand{\com}{[Co/M]}  
\newcommand{\nim}{[Ni/M]}  
\newcommand{\cum}{[Cu/M]}  
\newcommand{\cem}{[Ce/M]}  
\newcommand{\ndm}{[Nd/M]}  

\newcommand{\snr}{S/N}
\newcommand{\vk}{$V-K_{\rm s}$}
\newcommand{\jk}{$J-K_{\rm s}$}
\newcommand{\bprp}{BP$-$RP}
\newcommand{\vmicro}{$v_{\rm micro}$}
\newcommand{\vmacro}{$v_{\rm macro}$}
\newcommand{\vrad}{$v_{\rm rad}$}
\newcommand{\ebv}{$E(B-V)$}
\newcommand{\vsini}{$v \ \mathrm{sin} \ i$}
\newcommand{\mum}{$\mu$m}

\title{SDSS-V Milky Way Mapper (MWM): ASPCAP Stellar Parameters and Abundances in SDSS-V Data Release 19}

\author{
Szabolcs~M{\'e}sz{\'a}ros\altaffilmark{1,2}\orcid{0000-0001-8237-5209}, 
Paula~Jofr{\'e}\altaffilmark{3}, 
Jennifer~A.~Johnson\altaffilmark{4}, 
Jonathan~C.~Bird\altaffilmark{5}\orcid{0000-0001-5838-5212}, 
Jo~Bovy\altaffilmark{6}\orcid{0000-0001-6855-442X},
Andrew~R.~Casey\altaffilmark{7}, 
Julio~Chaname\altaffilmark{8}\orcid{0000-0003-2481-4546} 
Katia~Cunha\altaffilmark{9, 10}\orcid{0000-0001-6476-0576}, 
Nathan~De~Lee\altaffilmark{11}, 
Peter~Frinchaboy\altaffilmark{12}, 
Guillaume~Guiglion\altaffilmark{13,14,15}, 
Viola~Heged\H{u}s\altaffilmark{1,2}\orcid{0000-0001-7699-1902},
Alex~P.~Ji\altaffilmark{16,17}, 
Juna~A.~Kollmeier\altaffilmark{18, 19, 20}\orcid{0000-0001-9852-1610}, 
Melissa~K.~Ness\altaffilmark{21, 22},
Jonah~Otto\altaffilmark{12}, 
Marc~H.~Pinsonneault\altaffilmark{4}\orcid{0000-0002-7549-7766}, 
Alexandre~Roman-Lopes\altaffilmark{23}\orcid{0000-0002-1379-4204}, 
Andrew~Saydjari\altaffilmark{24}\orcid{0000-0002-6561-9002}
Amaya~Sinha\altaffilmark{25}, 
Ying-Yi~Song\altaffilmark{6, 26},
Guy~S.~Stringfellow\altaffilmark{27}, 
Keivan~G.~Stassun\altaffilmark{28}\orcid{0000-0002-3481-9052}, 
Jamie~Tayar\altaffilmark{29}\orcid{0000-0002-4818-7885}, 
Andrew~Tkachenko\altaffilmark{30}\orcid{0000-0003-0842-2374}, 
Marica~Valentini\altaffilmark{31}, 
Zachary~Way\altaffilmark{32}\orcid{0000-0003-0179-9662}, 
Jörg~Weingrill\altaffilmark{31}
}

\altaffiltext{1}{ELTE E\"otv\"os Lor\'and University, Gothard Astrophysical Observatory, 9700 Szombathely, Szent Imre H. st. 112, Hungary}
\altaffiltext{2}{MTA-ELTE Lend{\"u}let "Momentum" Milky Way Research Group, Hungary}
\altaffiltext{3}{Instituto de Estudios Astrof{\'i}sicos, Facultad de Ingenier{\'i}a y Ciencias, Universidad Diego Portales, Santiago de Chile}
\altaffiltext{4}{Department of Astronomy and Center for Cosmology and AstroParticle Physics, The Ohio State University, Columbus, OH 43210, USA}
\altaffiltext{5}{Department of Physics \& Astronomy, Vanderbilt University, 2301 Vanderbilt Place, Nashville, TN 37235, USA}
\altaffiltext{6}{Department of Astronomy and Astrophysics, University of Toronto, 50 St. George Street, Toronto, Ontario M5S 3H4, Canada ORCID iD}
\altaffiltext{7}{School of Physics \& Astronomy, Monash University, Wellington Road, Clayton 3800, Victoria, Australia}
\altaffiltext{8}{Instituto de Astrofísica, Pontificia Universidad Católica de Chile, Av. Vicuña Mackenna 4860, 782-0436 Macul, Santiago, Chile} 
\altaffiltext{9}{Steward Observatory, University of Arizona, 933 North Cherry Avenue, Tucson, AZ 85721-0065, USA}
\altaffiltext{10}{Observatório Nacional/MCTIC, R. Gen. José Cristino, 77,  20921-400, Rio de Janeiro, Brazil}
\altaffiltext{11}{Department of Physics, Geology, and Engineering Technology, Northern Kentucky University, Highland Heights, KY 41099, USA}
\altaffiltext{12}{Department of Physics and Astronomy, Texas Christian University, Fort Worth, TX 76129, USA}
\altaffiltext{13}{Zentrum f\"ur Astronomie der Universit\"at Heidelberg, Landessternwarte, K\"onigstuhl 12, 69117 Heidelberg, Germany}
\altaffiltext{14}{Max Planck Institute for Astronomy, K\"onigstuhl 17, 69117, Heidelberg, Germany}
\altaffiltext{15}{Leibniz-Institut f{\"u}r Astrophysik Potsdam (AIP), An der Sternwarte 16, 14482 Potsdam, Germany}
\hackaltaffiltext{16}{Department of Astronomy \& Astrophysics, University of Chicago, 5640 S Ellis Avenue, Chicago, IL 60637, USA}
\hackaltaffiltext{17}{Kavli Institute for Cosmological Physics, University of Chicago, Chicago, IL 60637, USA}
\hackaltaffiltext{18}{The Observatories of the Carnegie Institution for Science, 813 Santa Barbara Street, Pasadena, CA 91101, USA}
\hackaltaffiltext{19}{Canadian Institute for Theoretical Astrophysics, University of Toronto, Toronto, ON M5S-98H, Canada}
\hackaltaffiltext{20}{Canadian Institute for Advanced Research, 661 University Avenue, Suite 505, Toronto, ON M5G 1M1 Canada}
\hackaltaffiltext{21}{Department of Astronomy, Columbia University, Pupin Physics Laboratories, New York, NY 10027, USA}
\hackaltaffiltext{22}{Center for Computational Astrophysics, Flatiron Institute, 162 Fifth Avenue, New York, NY 10010, USA}
\hackaltaffiltext{23}{Department of Astronomy, Universidad de La Serena, Av. Raul Bitran \#1302, La Serena, Chile}
\hackaltaffiltext{24}{Department of Astrophysical Sciences, Princeton University, Princeton, NJ 08544, USA}
\hackaltaffiltext{25}{Department of Physics \& Astronomy, University of Utah, 115 South 1400 East, Salt Lake City, UT 84112, USA}
\hackaltaffiltext{26}{Dunlap Institute for Astronomy \& Astrophysics, University of Toronto, 50 St. George Street, Toronto, Ontario, M5S 3H4, Canada}
\hackaltaffiltext{27}{Center for Astrophysics and Space Astronomy, Department of Astrophysical and Planetary Sciences, University of Colorado, Boulder, CO 80309, USA}
\hackaltaffiltext{28}{Department of Physics and Astronomy, Vanderbilt University, Nashville, TN 37235, USA}
\hackaltaffiltext{29}{Department of Astronomy, University of Florida, Bryant Space Science Center, Stadium Road, Gainesville, FL 32611, USA}
\hackaltaffiltext{30}{Institute of Astronomy, KU Leuven, Celestijnenlaan 200D, 3001, Leuven, Belgium}
\hackaltaffiltext{31}{Leibniz-Institut für Astrophysik Potsdam (AIP), An der Sternwarte 16, 14482, Potsdam, Germany}
\hackaltaffiltext{32}{Department of Physics and Astronomy, Georgia State University, Atlanta, GA 30303, USA}


\begin{abstract}


The goal of this paper is to describe the science verification of Milky Way Mapper (MWM) APOGEE Stellar Parameter and Chemical Abundances Pipeline (ASPCAP) data products published in Data Release 19 (DR19) of the fifth phase of the Sloan Digital Sky Survey (SDSS-V). We compare MWM ASPCAP atmospheric parameters \teff, \logg, 24 abundances of 21 elements (carbon, nitrogen, and oxygen have multiple sources for deriving their abundance values) and their uncertainties determined from Apache Point Observatory Galactic Evolution Experiment (APOGEE) spectrograph spectra with those of the literature and evaluate their accuracy and precision. We also test the zero-point calibration of the \vrad \ derived by the APOGEE Data Reduction Pipeline. This data release contains ASPCAP parameters for 964,989 stars, including all APOGEE-2 targets expanded with new observations of 336,511 stars from the Apache Point Observatory observed until 2023 July 4. Overall, the new \teff \ values show excellent agreement with the IRFM scale, while the surface gravities exhibit slight systematic offsets compared to asteroseisimic gravities. The estimated precision of \teff \ is between 50 and 70~K for giants and 70$-$100~K for dwarfs, while surface gravities are measured with a precision of 0.07$-$0.09~dex for giants. We achieve an estimated precision of 0.02$-$0.04~dex for multiple elements, including metallicity, $\alpha$, Mg, and Si, while the precision of at least 10 elements is better than 0.1~dex.

\end{abstract}


\section{Introduction}

The information on the chemical composition of stars encoded in their spectra plays a central role in modern astrophysics. It allows us to probe the origins of various chemical elements, when and where they were produced and how they relate to the formation and evolution of galaxies, stars, and planets. One of the most intriguing questions of astronomy is how our Galaxy and its subsystems formed and evolved during its nearly 13 billion year lifetime. Answering this question became possible as a result of large spectroscopic sky surveys that observed the various subsystems of the Milky Way, namely the thin and thick disk, star clusters and the Galactic halo.

One such survey is the Milky Way Mapper (MWM; Johnson et al. in preparation), which is one of the three programs of the fifth generation of the Sloan Digital Sky Survey \citep[SDSS-V;][Kollmeier et al. 2025, in preparation]{2017arXiv171103234K}. This survey is obtaining high-resolution ($R$ = 22,500) stellar spectra in the $H$ band between 1.5 and 1.7 \mum \ of all stellar populations in the Milky Way using the APOGEE \citep{2019PASP..131e5001W}, and in the optical wavelengths with the low-resolution ($R$ = 1560 $-$ 2650) BOSS \citep{2013AJ....146...32S} spectrographs. Both are twin instruments located in the Northern hemisphere at the 2.5 m SDSS telescope \citep{2006AJ....131.2332G} and in the Southern Hemisphere at the 2.5 m Ir{\'e}n{\'e}e du Pont Telescope \citep{1973ApOpt..12.1430B} allowing for a scan of the entire sky from the ground.

A precursor of MWM was the Apache Point Observatory Galactic Evolution Experiment \citep[APOGEE; ][]{2017AJ....154...94M}, which was part of SDSS-III and SDSS-IV and used the same two high-resolution APOGEE spectrographs as MWM. The APOGEE Stellar Parameter and Chemical Abundances Pipeline \citep[ASPCAP; ][]{2016AJ....151..144G} was developed for the automated analysis of the observed APOGEE spectra to determine atmospheric parameters and chemical abundances by comparing observed spectra to libraries of theoretical spectra, using $\chi^2$ minimization in a multidimensional parameter space. ASPCAP was used to derive parameters and abundances for all APOGEE data releases from DR10$-$DR17 and was continuously updated over the years \citep{2013AJ....146..133M, 2015AJ....150..148H, 2018AJ....156..125H, 2018AJ....156..126J, 2020AJ....160..120J}. In its final data release, DR17 \citep{2022ApJS..259...35A} APOGEE published atmospheric parameters of nearly 650,000 individual stars. The observing plan of both APOGEE and MWM was put together to cover all of the stellar populations of the Milky Way so that the derived stellar properties allowed us to explore the properties of our Galaxy as a centerpiece of near-field cosmology. Indeed, the Milky Way is the only case study of the path from random initial fluctuations in the Universe to a mature galaxy, because only in the Milky Way can we observe and analyze individual stars that formed at all epochs of the cosmic timeline. 

By using ASPCAP data, several important discoveries have been made in the field of galactic chemical evolution. By studying the spatial distribution of $\alpha$ elements with \feh \ in different spatial zones of the Galactic disk, \citet{2014A&A...564A.115A, 2014ApJ...796...38N, 2015ApJ...808..132H} showed that the bimodal distribution of $\alpha$ elements exists over a large extent of the disk, even in the center of the Milky Way. The detailed chemical makeup of the Bulge has been discussed by \citet{2016PASA...33...22N}, \citet{2018ApJ...852...91G}, and \citet{2019ApJ...870..138Z} who analyzed the distribution of metallicity and 11 individual elements in the inner Galaxy. These observations of the bulge, disk, and halo combined through the uniform ASCPAP abundances put important constraints on galactic chemical evolution models aiming to explain how our Galaxy has formed in its entirety.

The APOGEE observations made it possible to study abundances of stars in the satellite galaxies of the Milky Way in detail \citep{2018Natur.563...85H, 2020ApJ...895...88N, 2021ApJ...923..172H}, whose chemical makeup may be affected by their interaction with our Galaxy. For example, \citet{2020ApJ...895...88N} argued that the [$\alpha$/Fe]$-$[Fe/H] distribution of the Large Magellanic Cloud (LMC) indicates a relatively recent starburst. They suggested that LMC originally formed in a lower-density environment than the Milky Way, making its way into the Milky Way's gravitational potential only recently. Using ASPCAP parameters and abundance trends of 11 elements, \citet{2021ApJ...923..172H} were able to study all of the massive satellite galaxies of the Milky Way: the Large and Small Magellanic Clouds (LMC/SMC), the Sagittarius Dwarf Galaxy (Sgr), Fornax (Fnx), and the now fully disrupted Gaia Sausage/Enceladus (GSE) system. They showed that the starburst in the SMC happened 3$-$4 Gyr before the burst in LMC. Evidence of secondary star-forming events was also found in Sgr and Fnx roughly 5$-$7 Gyr ago, but not in GSE, which appears to have had stronger initial star formation as compared to the other four galaxies. Comparing these chemical patterns with that of our Galaxy helps to better understand the overall evolution of the Local Group.

The scientific reach of the ASPCAP results is not limited to answering questions about the formation of our Galaxy and its immediate surroundings, but provides important data for understanding stellar evolution in general. For example, ASPCAP measurements of abundances of carbon and nitrogen made it possible to derive stellar ages for a larger number of stars. \citet{2016MNRAS.456.3655M} and \citet{2016ApJ...823..114N} derived empirical calibration using red giants with asteroseismic mass estimates from Kepler and ASPCAP \teff, \logg, \mh, and [C/N] values that allowed them to estimate stellar masses with uncertainties of about 14~\% and ages with rms errors of 40~\% for 52,000 and 70,000 stars, respectively. By combining Two Micron All Sky Survey (2MASS) photometry with stellar evolution models and ASPCAP stellar parameters and abundances, \citet{2014ApJ...790..127B} were able to differentiate a large sample of red-clump stars from their red giant counterparts, and by using the narrowness of the RC locus in color$-$metallicity$-$luminosity space they assigned distances to the stars with an accuracy of 5\%$-$10\%. The catalog was useful in studying the stellar distribution function of spatial location, kinematics, and abundances in the Milky Way disk. These studies only highlight the possibilities that may arise when MWM data is combined with other surveys, which has the potential to use the stellar velocities, ages, temperatures, luminosities, elemental abundances, etc., to tackle a huge variety of problems ranging from galaxy evolution, dark matter distribution, metal production, star formation, binary interaction, to exoplanet demography. 

\begin{figure*}                          
\centering
\includegraphics[width=7.1in,angle=0]{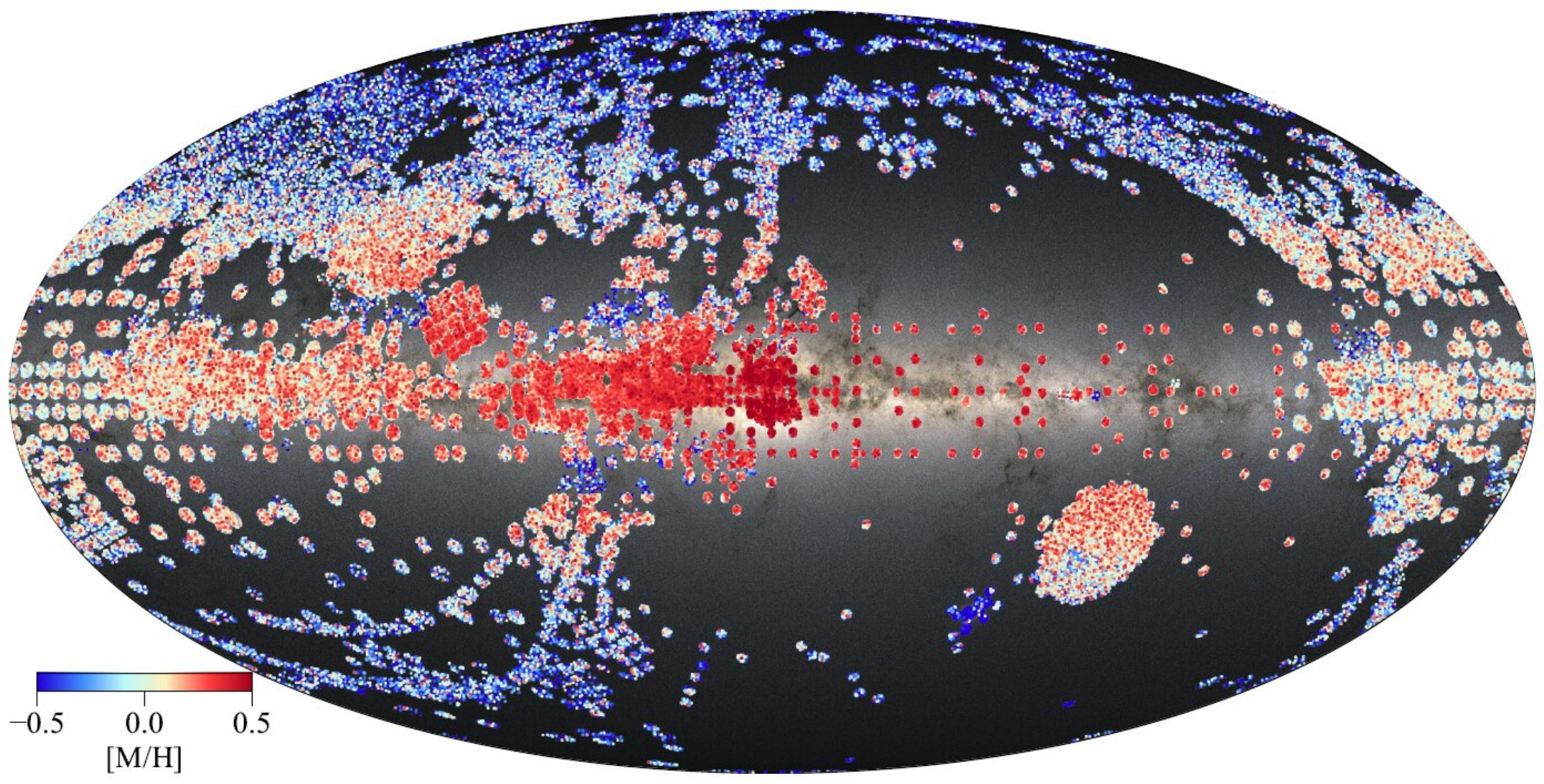}
\caption{Position of stars that have parameters and abundances below 8000~K published by Milky Way Mapper in DR19 color coded by \mh. Only stars with SNR $>$ 50 per pixel and no BAD flags that are farther than 1000~pc are shown here to reveal the metallicity gradient of our Galaxy.}
\label{mwmet}
\end{figure*}

While ASPCAP has been a very successful pipeline for APOGEE, the expanded observations of MWM in the Hertzsprung–Russell diagram  required extra methodologies to be implemented that are appropriate for the analysis of white dwarfs, hot massive stars, and young stellar objects. This, combined with the fact the MWM also needs to analyze the optical spectra of the BOSS spectrograph, made it necessary to expand the methods used to derive those parameters from stellar spectra observed by MWM. Astra is the analysis framework for MWM (Casey et al. 2025, in preparation). Its purpose is to manage the analysis of reduced data products from SDSS-V to provide atmospheric parameters and abundances for upcoming data releases. Astra is capable of running multiple algorithms and methods that have been developed for such purposes, including but not limited to: ASPCAP \citep{2016AJ....151..144G}, APOGEE Net \citep{2020AJ....159..182O, 2022AJ....163..152S}, SLAM \citep{2020ApJS..246....9Z}, Payne \citep{2019ApJ...879...69T}, and Cannon \citep{2015ApJ...808...16N}. In this data release, all prior APOGEE data were reanalyzed by Astra. New SDSS observations from APO of 336,511 targets have been reduced and processed for DR19, extending the number of observed stars to 964,989.

The focus of this paper is to describe the accuracy, precision, and science verification of parameters derived by ASPCAP and published in the 19th data release of SDSS. These parameters include radial velocities, atmospheric parameters, and the elemental abundances. Section 2 gives a short overall description of how ASPCAP works and describes the quality of the radial velocities. The accuracy and precision of ASCPAP main parameters are discussed in Section 3, while the quality of ASPCAP abundances is detailed in Sections 4 and 5.


\section{ASPCAP Overview}

The primary pipeline for the derivation of stellar parameters for the APOGEE survey was ASPCAP \citep{2016AJ....151..144G} and it remains a critical tool for the analysis of spectra of FGKM stars in the MWM survey. ASPCAP does a pseudo-continuum-normalization, and then its core code, FERRE \citep{2006ApJ...636..804A} interpolates a precomputed grid of synthetic spectra to find the best-fitting stellar parameters that describe an observed spectrum using a $\chi^2$ minimization algorithm. After a coarse-characterization of spectra to find which subgrid to use for a detailed analysis, it first performs a global simultaneous fit of eight stellar parameters, which include effective temperature (\teff), surface gravity (\logg), metallicity \mh, microturbulent velocity (\vmicro), macroturbulent velocity (\vmacro) for giants or rotational velocity (\vsini) for dwarfs, and abundances of $\alpha$ elements (\am), carbon (\cm), and nitrogen (\nm). These are considered the main parameters in ASPCAP. 

The last three are necessary in the global fit because the $\alpha$ elements are important electron donors in the atmosphere of FGKM stars, and with the formation of CO, they have a significant effect on the equation of state, making it essential to take them into account when fitting the full spectrum before the abundances of individual elements are determined. OH, CO, and CN molecular absorption lines cover a large fraction of the H-band spectrum. In addition, some of the MWM targets might have enhanced carbon abundances, which significantly changes the atmospheric structure compared to the solar case, justifying the use of carbon-enhanced model atmospheres for the spectral synthesis \citep{2012AJ....144..120M}. In the second step, chemical abundances are determined in wavelength windows sensitive to each specific element with the main parameters fixed. For DR19, abundances of 21 different elements are determined by ASPCAP. 

\begin{figure*}                          
\centering
\includegraphics[width=7.1in,angle=0]{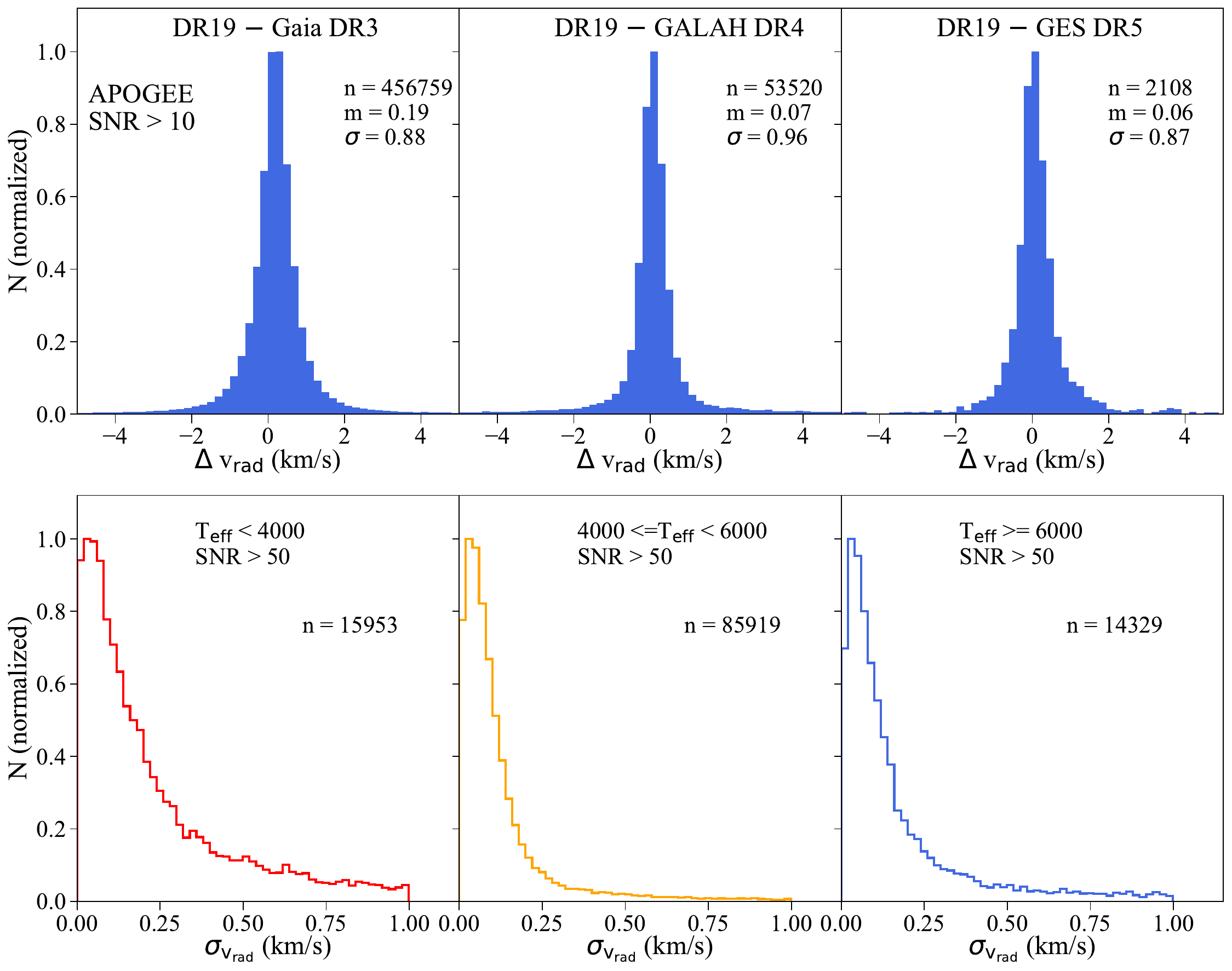}
\caption{Radial velocity (top panels) published in DR19 compared with Gaia DR3 \citep{2023A&A...674A..33G}, GALAH DR4 \citep{2024arXiv240919858B}, and GES DR5 \citep{2022A&A...666A.121R}. The scatter of radial velocities as a function of \teff \ is shown in the bottom panels.}
\label{radvelcomp}
\end{figure*}

DR19 makes use of the spectral synthetic grids developed for APOGEE in DR17 \citep{2022ApJS..259...35A}. These spectral grids have been constructed using the Synspec \citep{2021arXiv210402829H} spectral synthesis code, which incorporates both local thermodynamic equilibrium (LTE) and non-LTE (NLTE) calculations. Although NLTE level populations for Na, Mg, K, and Ca \citep{2020A&A...637A..80O} were created for APOGEE DR17, the abundances derived in DR19 are all using the LTE approximation. The syntheses are based on MARCS model atmospheres that were computed for APOGEE DR14 and DR16, as described in \citet{2018AJ....156..125H} and \citet{2020AJ....160..120J}. The atomic and molecular line list data used for synthesis is described by \citet{2016ApJ...833...81H}, \citet{2017ApJ...844..145C} and \citet{2021AJ....161..254S}, who adjusted the atomic data to fit the spectra of the Sun and Arcturus with the \citet{2007SSRv..130..105G} abundances. The abundances published in DR19 are closely tied to the \citet{2007SSRv..130..105G} solar abundance scale as both the model atmospheres and synthesis were calculated using this abundance table. 

The parameter space is so large that it is not possible to read the synthetic spectral database into memory in its entirety with one single grid. Thus, several spectral grids were constructed to properly derive parameters across the HR-diagram. The coverage of the different spectral grids is described in Section 4.1, and listed in Table 2 of \citet{2020AJ....160..120J}. 

While ASPCAP was run on all observed APOGEE spectra, here we focus on FGKM stars that were observed with the APOGEE spectrograph and whose temperature is lower than 8000~K. The choice of this temperature range is motivated by the fact that most abundances can only be derived below 7500$-$8000~K in the $H$ band spectra of FGKM type stars, making them the ideal tracers of stellar population studies of the Milky Way. There are 894,256 individual stars below this temperature limit. A subset of this ASPCAP sample is visualized in Galactocentric coordinates, color coded by metallicity for stars with no BAD flags, signal-to-noise ratio (\snr) $>$ 50 per pixel, and distance larger than 1000 pc to reveal the metallicity gradient of the Milky Way in Figure~\ref{mwmet}.

\subsection{Radial Velocity}

The radial velocities (\vrad) published in DR19 are barycentric velocities, not heliocentric ones. Their accuracy can be checked by comparing their values with the ones derived by other spectroscopic surveys. The comparison of APOGEE \vrad \ values with Gaia DR3 \citep{2023A&A...674A..33G}, GALAH DR4 \citep{2024arXiv240919858B} and Gaia-ESO DR5 \citep{2022A&A...666A.121R} can be seen in the top panels of Figure~\ref{radvelcomp}. The following parameter restrictions were applied when making these comparisons: the error (e\_v\_rad) and scatter of \vrad \ (std\_v\_rad) were both set to be lower than 1 \kms \ to throw out potential variable stars, and \snr \ was set to be higher than 10 in both samples. Spectra with \snr \ $<$ 10 were rejected because RV measurements from such low-quality data were proved to be unreliable.

\begin{deluxetable*}{lrrrrrccr}
\tabletypesize{\scriptsize}
\tablecaption{Overview of the reference samples used to examine the precision and accuracy of DR19 ASPCAP data.}
\tablewidth{0pt}
\tablehead{
\colhead{Reference Sample} & \colhead{\teff\ Range} & 
\colhead{\logg \ Range}  &
\colhead{[M/H] Range} & 
\colhead{N Stars} & 
\colhead{Purpose} & 
\colhead{In Text} &
\colhead{Comments}}
\startdata
Low $E(B-V)$ & [3900, 8000] & [$-$0.5, 5.0] & [$-$2.5, 0.6] & 223\,806 &  \teff \ validation & Sec.~\ref{sec:teff} & \ebv \ $<$ 0.02, \\
&  &  &  &  &  &  & and defined using \\
&  &  &  &  &  & & Gaia colors \\ \\

Gaia Benchmarks & [3900, 6800]  & [0.9, 5.0] & [-2.5, 0.6] & 55 & \teff, \logg, \mh & Sec.~\ref{sec:teff}, \ref{secgrav}, ~\ref{sec:metallicity} & \\ 
&  &  &  &  & accuracy and precision &  & \\ 
&  &  &  &  & assessment & \\ \\

APOKASC3 & [4200, 5300] & $<$ 3.8 & [$-$0.8, 0.5] & 22\,468 &  \logg \ validation & Sec.~\ref{secgrav} & Further refined using \\ 
 & & & & & & & APOKASC3 flags \\ \\

TESS & [4000, 5500] & $<$ 3.8 & [$-$1.0, 0.5] & 24\,136 &  \logg \ validation & Sec.~\ref{secgrav} &  \\ 
 & & & & & & & \\ \\

Open Clusters & [4000, 6000] &  $<$ 3.8 & [$-$0.4, 0.5] & 358 & \mh \ and \xm \ accuracy & Sec.~\ref{sec:metallicity}, \ref{abuteffsec}, ~\ref{elements} & \\ 
 & & & & & and precision assessment & Tab.~\ref{corrcoeff}, ~\ref{abuerrorstab}, Eq.~\ref{teffabueq} & \\ \\ 

Globular Clusters & [4000, 6000] & $<$ 3.8 & [$-$2.5, $-$0.6] & 1\,344 & [M/H] accuracy and & Sec.~\ref{sec:metallicity} & \\ 
 & & & & & precision assessment & & \\ \\

MWM-GALAH DR4 & [3500, 8000] & [0.0, 5.0] & [$-$2.5, 0.6] & 63\,136 & \mh \ and \xm \ & Sec.~\ref{sec:metallicity}, ~\ref{elements} & GALAH S/N $>$ 50 \\
common sample & & & & & accuracy assessment & & \\ \\

MWM-Gaia-ESO DR5 & [3600, 7000] & [0.0, 5.0] & [$-$2.5, 0.6] & 2\,660 &  \mh \ and \xm \  & Sec.~\ref{sec:metallicity} \\ 
common sample & & & & & accuracy assessment  & \\ \\

Solar Neighborhood & [3200, 8000] & [0.0, 5.0] & [$-$0.05, 0.05] & 42\,376 & \xh \ calibration and & Sec.~\ref{accuabusec}, ~\ref{elements} & no flags \\
 & & & & & \xm \ precision assessment & Tab.~\ref{zeropoint}, ~\ref{abuerrorstab} & \\ \\ 

Wide Binaries & [4000, 6000] & [3.0, 5.0] & [$-$1.0, 0.5] & 696 & \xm \ precision assessment & Sec.~\ref{precabu}, ~\ref{elements} & \\ 
 & & & & &  & Tab.~\ref{abuerrorstab} & \\ \\ 

\enddata 
\tablecomments{All these samples have stars with S/N $>$ 50 and no BAD flags.}
\label{tab:validation-set}
\end{deluxetable*}

The best agreement is achieved with Gaia-ESO DR5 for which the median difference is 0.06~\kms, though the common sample contains only 2108 stars, 1$-$2 orders of magnitude fewer than in case of Gaia and GALAH. Similar discrepancies can be observed in the DR19$-$Gaia DR3 and GALAH DR4 common sample at 0.19 and 0.07~\kms, respectively. The standard deviation of differences is quite similar in all three samples ranging between 0.87 and 0.96~\kms. A few objects that have differences larger than 5~\kms \ are not taken into account when calculating the scatter. These may be variable stars that remained in the sample even after restricting the error and scatter of the individual \vrad \ measurements.

The precision of radial velocities can be estimated by examining the scatter of derived radial velocities in stars with multiple visits. The histogram of this scatter for the APOGEE spectrograph is shown in the bottom panels of Figure~\ref{radvelcomp} as a function of \teff. The precision of the DR19 \vrad \ values is excellent, peaking at around 70~\ms \ when considering spectra with \snr \ $>$ 50 and temperatures below 8000~K. Partway through SDSS-V, APOGEE was upgraded to include a Fabry-Perot interferometer standard, back pressure regulator and octagonal fibers, all of which improved the radial velocity stability. As DR19 includes many DR17 spectra reduced without these improvements, we note the overall radial velocity performance is better than 70 m/s for stars taken in SDSS-V after the upgrades (Kollmeier et al 2025 in preparation). \citet{2024arXiv240807126S} re-reduced the APOGEE-1 and -2 data with the apMADGICS package. They showed that the RV precision reached 30 m/s in SDSS-IV if the RV comparisons were restricted to RVs measured in the same fiber and 47 m/s measured in different fibers for stars with \snr \ $>$ 120. The hardware and software improvements in SDSS-V counteract the fiber-to-fiber differences.

The precision is somewhat lower, between 70 and 100~\ms \ for the cool stars below 4000~K, which may be the result of the difficulty of getting accurate atmospheric parameters for the radial velocity templates. At these temperatures, the strength of atomic and molecular lines in the $H$ band are very sensitive to the temperature, and a template with incorrect \teff \ may look different enough from the observed spectra to produce higher errors than at higher temperatures. However, the measured precision and accuracy of the APOGEE radial velocities vastly exceed the limit necessary to measure accurate atmospheric parameters and abundances.

\begin{figure*}                         
\centering
\includegraphics[width=7.1in,angle=0]{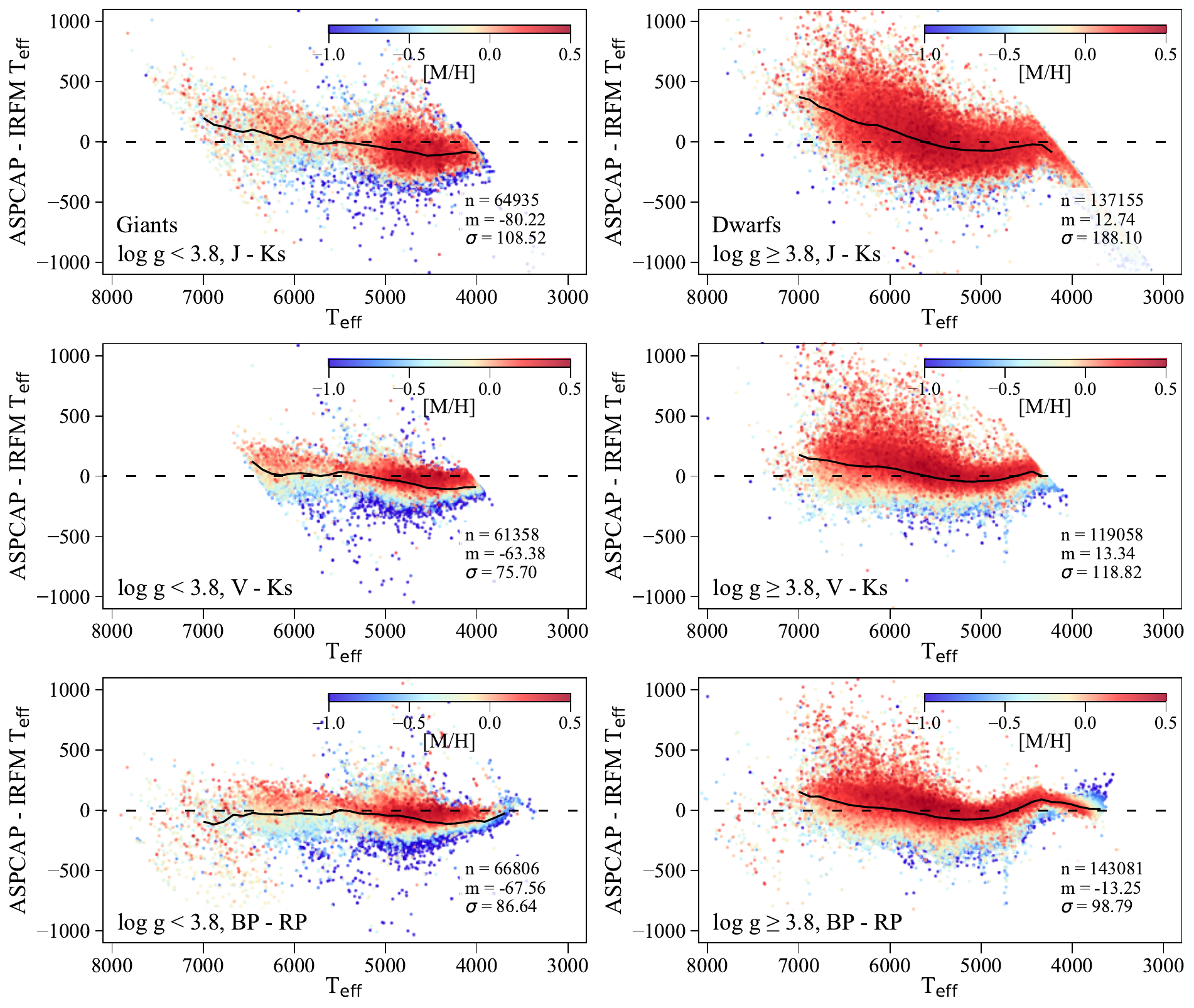}
\caption{Differences of DR19 ASPCAP spectroscopic \teff \ and IRFM temperatures as a function of ASPCAP \teff \ color coded by metallicity. The left panels show the giants, the right panels the dwarfs. Photometric \jk \ and \vk \ temperatures from \citet{2009A&A...497..497G} can be seen in the top and middle panels, respectively. The bottom panel shows the differences compared to IRFM temperatures calculated from BP$-$RP colors \citep{2021MNRAS.507.2684C}. The solid black line shows the running median value.
}
\label{fig:teffcal}
\end{figure*}


\section{ASPCAP Main Atmospheric Parameters}

Atmospheric parameters and abundances derived from stellar spectra are inherently inaccurate, because no theoretical models can perfectly describe the observed spectrum. In order to provide parameters and chemical abundances that are as accurate as possible, ASPCAP in DR19 emulates previous SDSS Data Releases by using a variety of reference sets to evaluate the accuracy and precision of parameters. The reference sets considered here in DR19 are summarized in Table~\ref{tab:validation-set}, where we also indicate their purpose. Details about how each of them is assembled and used are extensively explained throughout the subsequent sections. As extensively discussed in \cite{2019ARA&A..57..571J}, providing accurate parameters and abundances makes it easier to complement the data with other stellar surveys of the Milky Way, as well as to improve theoretical models of stellar, chemical, and galaxy evolution in general. By providing precise parameters and abundances, the data serves to study robust trends and to explore the various patterns in chemistry over space and time. 

The complicated relationship between the main parameters may lead to correlations and offsets as a function of the main parameters themselves. Therefore, similarly to some of the previous APOGEE data releases, \teff \ and \logg \ values were calibrated to independent measurements, as detailed in Casey et al. (2025, in preparation). Both the raw and calibrated parameters have their own named columns in DR19. The calibrated values are stored in teff and logg, while their original values can be found in raw$\_$teff and raw$\_$logg. Metallicity is not calibrated in DR19, and it is stored as raw$\_$m$\_$h$\_$global. In this section, we discuss the accuracy and precision of the raw values of \teff, \logg, and \mh, because these are the values used to derive the abundances of individual elements (see Section~\ref{sec:abundances} for details). 


\subsection{Effective Temperature}
\label{sec:teff}


\begin{figure*}                          
\centering
\includegraphics[width=7.1in,angle=0]{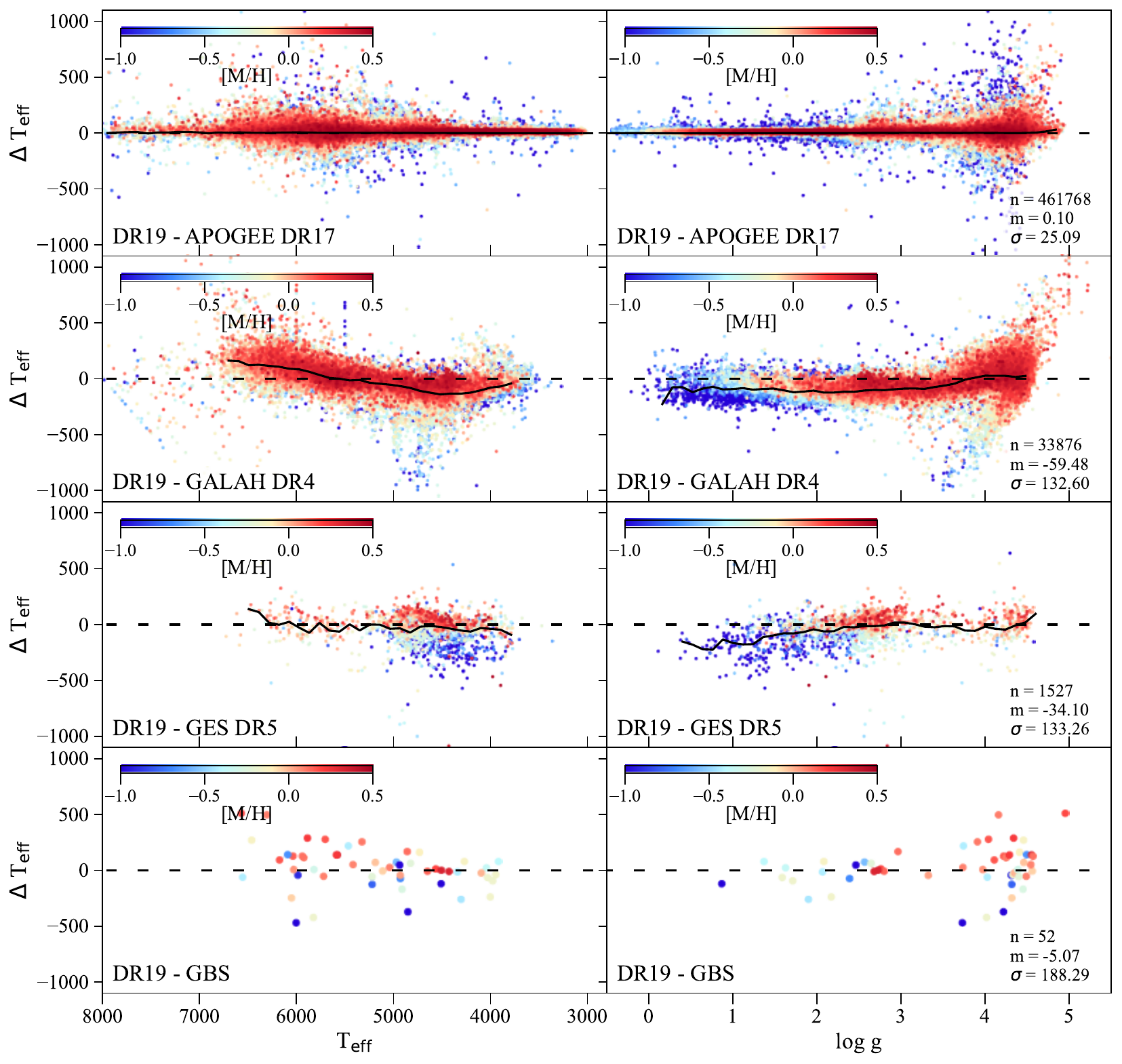}
\caption{Differences of DR19 ASPCAP spectroscopic raw \teff \ and APOGEE DR17 \citep{2022ApJS..259...35A}, GALAH DR4 \citep{2024arXiv240919858B}, Gaia-ESO DR5 \citep{2022A&A...666A.121R} and GBS \citep{2024A&A...682A.145S} temperatures as a function of ASPCAP raw \teff \ color coded by raw metallicity. The solid black line shows the running median value.}
\label{fig:teffcomp}
\end{figure*}

Effective temperatures are first compared to three different photometric scales based on the infrared flux method (IRFM): from \jk \ and \vk \ colors of \citet[][hereafter GHB]{2009A&A...497..497G}, and \bprp \ of \citet{2021MNRAS.507.2684C}. IRFM calculates temperatures using the Stefan-Boltzmann law, but relies on model atmospheres rather than on a direct measure of the angular diameter; thus, IRFM temperatures are semi-directly tied to the absolute temperature scale. Because GHB defined two separate temperature scales for giants and dwarfs, two separate equations are used for comparisons, where the giants were defined as stars having \logg \ $<$ 3.8. The photometric \teff \ equation used by \citet{2021MNRAS.507.2684C} applies to all surface gravities. However, we divided them into giants and dwarfs using the definition above to be able to compare all three IRFM scales with the ASPCAP values.

To establish the basis of our comparison set, we selected stars that have \ebv \ $<$ 0.02, \snr \ $>$ 50, and flag$\_$bad = false. \ebv \ values were taken from the column named ebv in the DR19 ASPCAP file, which was specially calculated for DR19 as detailed by Casey et al. (2025, in preparation). The overall parameter range of the low \ebv \ sample and number of stars is listed in Table~\ref{tab:validation-set}. The difference between the ASPCAP raw \teff \ and the GHB temperature from the \jk \ and \vk \ colors can be seen in the top and middle panel of Figure~\ref{fig:teffcal}, respectively, while comparison between ASPCAP raw \teff \ and the photometric temperature from \bprp \ color is shown in the bottom panel of Figure~\ref{fig:teffcal}. The agreement between all temperature scales can be considered very good; however, small systematic offsets are present both as a function of \teff, and \mh. The temperature dependence is visible for both the main sequence stars (right panels of Figure~\ref{fig:teffcal}) and giant stars (left panels of Figure~\ref{fig:teffcal}), but the effect is stronger for dwarfs, as ASPCAP temperatures above \teff \ $>$ 5500~K are warmer than the IRFM ones and also have larger scatter when compared to all three IRFM temperatures. The other obvious correlation can be seen as a function of metallicity especially for giants, in which metal-poor stars have lower temperatures down to about 200$-$400~K than the IRFM scales. 

In addition to these minor correlations, there is also a systematic offset between ASPCAP and the IRFM scale for giants. The median difference below for \logg \ $<$ 3.8 varies between $-$63.4 and $-$80.2~K, but for dwarfs, the agreement is excellent with offsets between $-$13.3 and 13.3~K. The GHB scales do not allow for a comparison below about 4200~K for main-sequence stars, but the temperature scale from the \bprp \ color goes down to about 3800$-$4000~K. To further assess the accuracy of the temperatures, the comparison between our raw \teff\ values and the ones from APOGEE DR17 \citep{2022ApJS..259...35A}, GALAH DR4 \citep{2024arXiv240919858B}, Gaia-ESO DR5 \citep{2022A&A...666A.121R}, and the Gaia FGK Benchmark stars \citep[][hereafter GBS; see Table~\ref{tab:validation-set}]{2024A&A...682A.145S} can be found in Figure~\ref{fig:teffcomp}. While the agreement with APOGEE DR17 temperatures is excellent, there are small systematic discrepancies observed with relative to GALAH DR4 and Gaia-ESO DR5. The GBS temperatures are derived using the Stefan-Boltzmann relation with direct measurements of the angular diameter, precise parallaxes from Gaia, and accurate bolometric fluxes from photometry. This makes the GBS an excellent reference for fundamental temperatures. The agreement between the two temperature scales is excellent, as the median difference is only $-$5.07~K, though the small number of stars in common does not allow us to uncover more complex relationships, as found in the comparison with GHB temperatures.

The uncertainty of \teff \ calculated by Astra is stored in e\_teff. Its median value in the entire sample is only 6.2~K. The precision of temperature can also be estimated using the standard deviation around mean values of the discrepancy between the spectroscopic, photometric, and the GBS temperatures. The average GBS temperature uncertainty in this sample is 53.3~K, though this is based on only 52 stars. The scatter around the differences of ASPCAP and IRFM temperatures varies between 75.7 and 108.5~K for giants and 98.8~K and 188.1~K for dwarfs. 

The wider wavelength coverage of \vk\ compared to J$-$K$_{\rm s}$, and the higher precision of the $V$ magnitudes than the J magnitudes provides a more precise relationship between color and \teff. This also explains why the scatter of differences is larger in the top two panels of Figure~\ref{fig:teffcal} than the two middle ones. The $V$ magnitudes are taken from the synthetic photometry database of Gaia DR3 \citep{2023A&A...674A..33G}, which used BP/RP spectra to determine their values. The higher scatter seen in the \jk \ comparison is probably the result of the relatively higher errors of the 2MASS photometry compared to Gaia; thus, the \vk \ and \bprp \ colors give a more precise estimate of the precision of ASPCAP \teff. Considering the relatively small errors of the photometry, the uncertainty of the reddening (whose values are limited to be smaller than 0.02), and the calibration of the IRFM scale, we estimate that the overall uncertainty of ASPCAP temperatures is around 50$-$70~K for giants, and 70$-$100~K for main-sequence stars. Thus, the errors reported by Astra are probably underestimated in most cases.

\begin{figure}                          
\centering
\includegraphics[width=3.42in,angle=0]{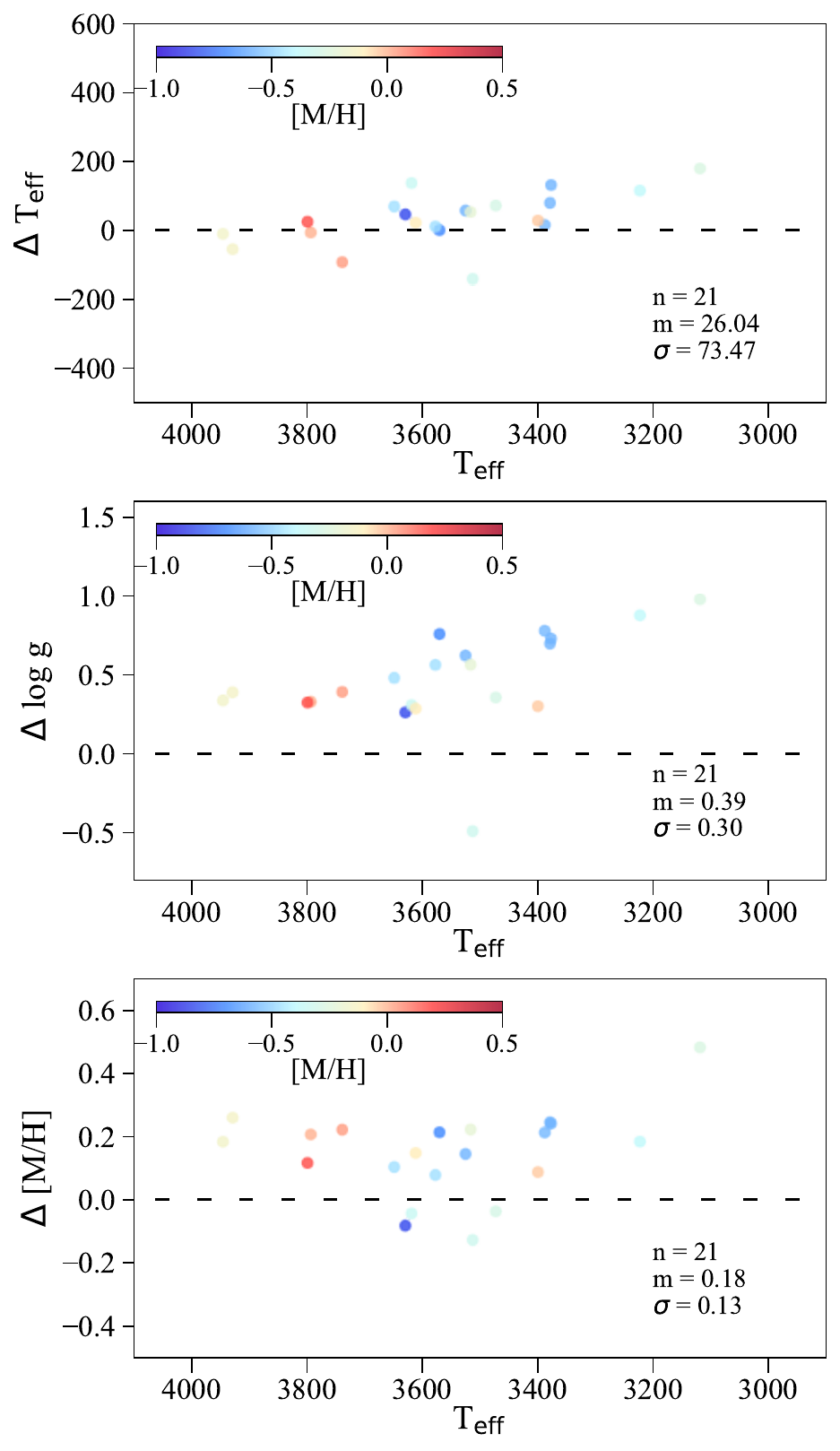}
\caption{Main atmospheric parameters comparison with those of derived by \citet{2022ApJ...927..123S} as a function of ASPCAP raw \teff values. Top panel: difference of temperatures from \citet{2022ApJ...927..123S} and ASPCAP raw \teff \ values; middle panel: difference of surface gravities from \citet{2022ApJ...927..123S} and ASPCAP raw \logg \ values; bottom panel: difference of metallicities from \citet{2022ApJ...927..123S} and ASPCAP raw \mh \ values.}
\label{mdwarfs}
\end{figure}

\subsubsection{M Dwarfs}
\label{mdwarfteff}

Interestingly, the ASPCAP temperatures of main-sequence stars below 4500~K, cool K dwarfs and M dwarfs, becomes hotter than the \bprp \ scale by about 100$-$200~K at about \mh \ = $-$1~dex (see bottom-right panel of Figure~\ref{fig:teffcal}), but the two temperature scales agree well near solar metallicity. \citet{2022ApJ...927..123S} derived main parameters and individual abundances independently from ASPCAP of 21 M benchmark dwarfs observed by APOGEE-2 using samples of H$_{\rm 2}$O and OH lines. The comparison of their temperatures with ASPCAP DR19 raw ones shows a similar effect, good agreement at solar metallicity, but slight offsets about 100$-$200~K at lower metallicities (top panel of Figure~\ref{mdwarfs}). It is possible that this signals a problem with accurately determining the atmospheric parameters of metal-poor red dwarf stars below 4500~K (see Sections~\ref{mdwarflogg} and~\ref{mdwarfmet} for discussion of \logg \ and \mh \ values of red dwarfs, respectively).


\subsection{Surface Gravity}
\label{secgrav}

\begin{figure}                         
\centering
\includegraphics[width=3.42in,angle=0]{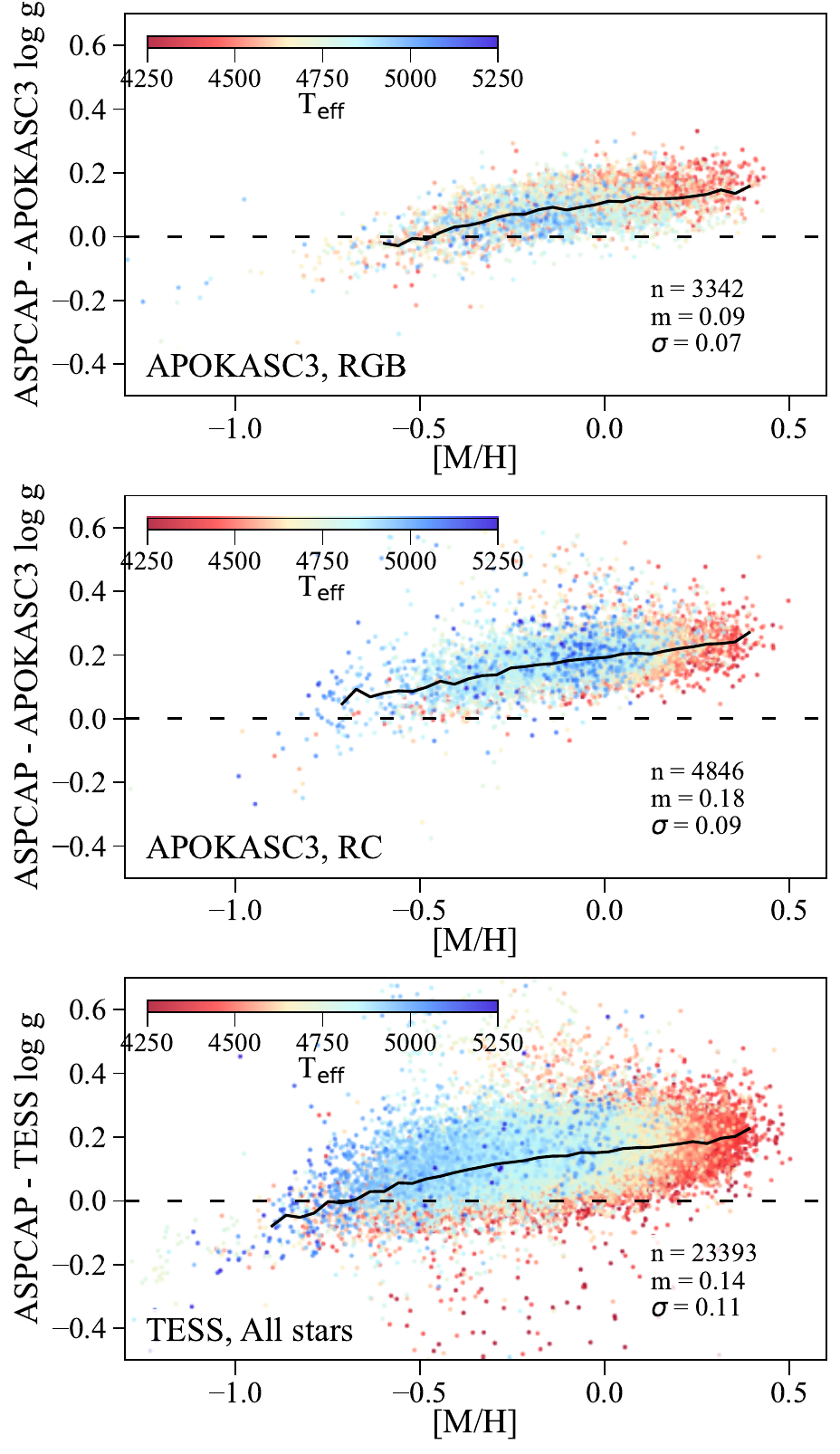}
\caption{Differences of DR19 ASPCAP spectroscopic and astroseismic \logg \ as a function of ASPCAP metalliciy and raw surface gravity. The top panel shows the RGB, the middle panel the RC stars from the APOKASC3 sample, and the bottom panel shows all TESS stars.}
\label{fig:loggcal}
\end{figure}

The spectral features in APOGEE spectra are less sensitive to surface gravity than the other main parameters; thus, achieving accurate and precise measurements of \logg\ is often challenging. Acknowledging this, a dedicated effort has been invested over several years in SDSS to take APOGEE spectra of Kepler, K2, and TESS stars \citep{2014ApJS..215...19P, 2018ApJS..239...32P, 2024AJ....167...50S, 2023RNAAS...7..148T}. This is because very accurate surface gravities can be derived from the oscillations of pulsating variable stars through asteroseismic analysis. The latest of such efforts, namely the APOKASC3 catalog, was compiled by \citet{2024arXiv241000102P} from observation of variable stars across the red giant branch (RGB), red clump (RC), and asymptotic giant branch \citep{2024arXiv241103101V} with turnoff and-main sequence stars also available in a separate catalog \citep{2017ApJS..233...23S}. The APOKASC3 catalog contains masses, radii, ages, and surface gravities tied to a fundamental scale. In particular, the typical uncertainties of their \logg \ values are of the order of 0.01$-$0.02~dex. A similar catalog was put together from the observations of the TESS planet hunting mission \citep{2015JATIS...1a4003R} using seismic results from \citet{2021ApJ...919..131H} and a procedure similar to \citet{2023RNAAS...7..148T}. The MWM-TESS common sample contains 24,136 observations of 19,582 unique stars with similar precision of surface gravities to the APOKASC3 catalog.

\begin{figure}                          
\centering
\includegraphics[width=3.42in,angle=0]{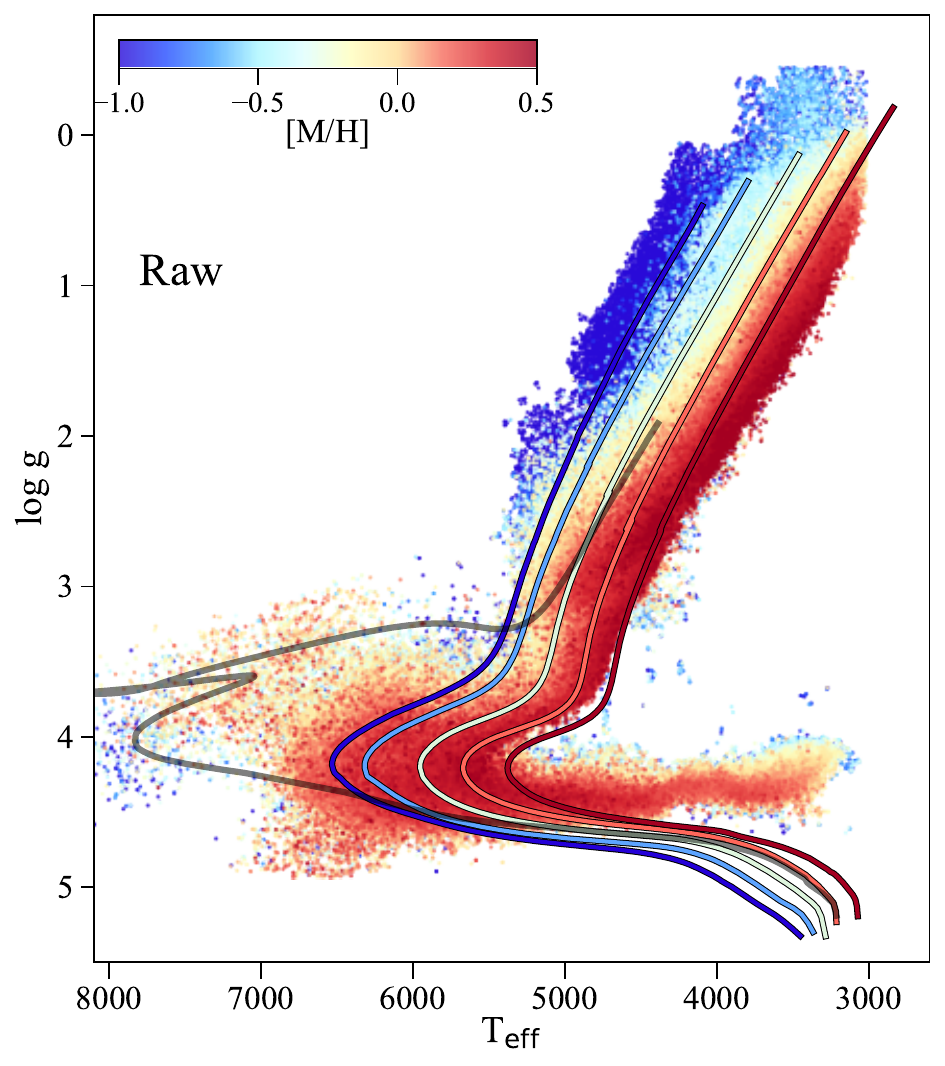}
\caption{Kiel diagram showing raw \teff \ and \logg \ values. Dartmouth isochrones with ages of 12 Gyr, and \mh \ = $-$1.5 to 0.5 dex are overplotted. The light-gray line shows a 1 Gyr old solar-metallicity isochrone. Note that the \logg \ discrepancy between the isochrones and ASPCAP increases below 4500~K in the main sequence.}
\label{fig:tefflogg}
\end{figure}

\begin{figure*}                          
\centering
\includegraphics[width=7.1in,angle=0]{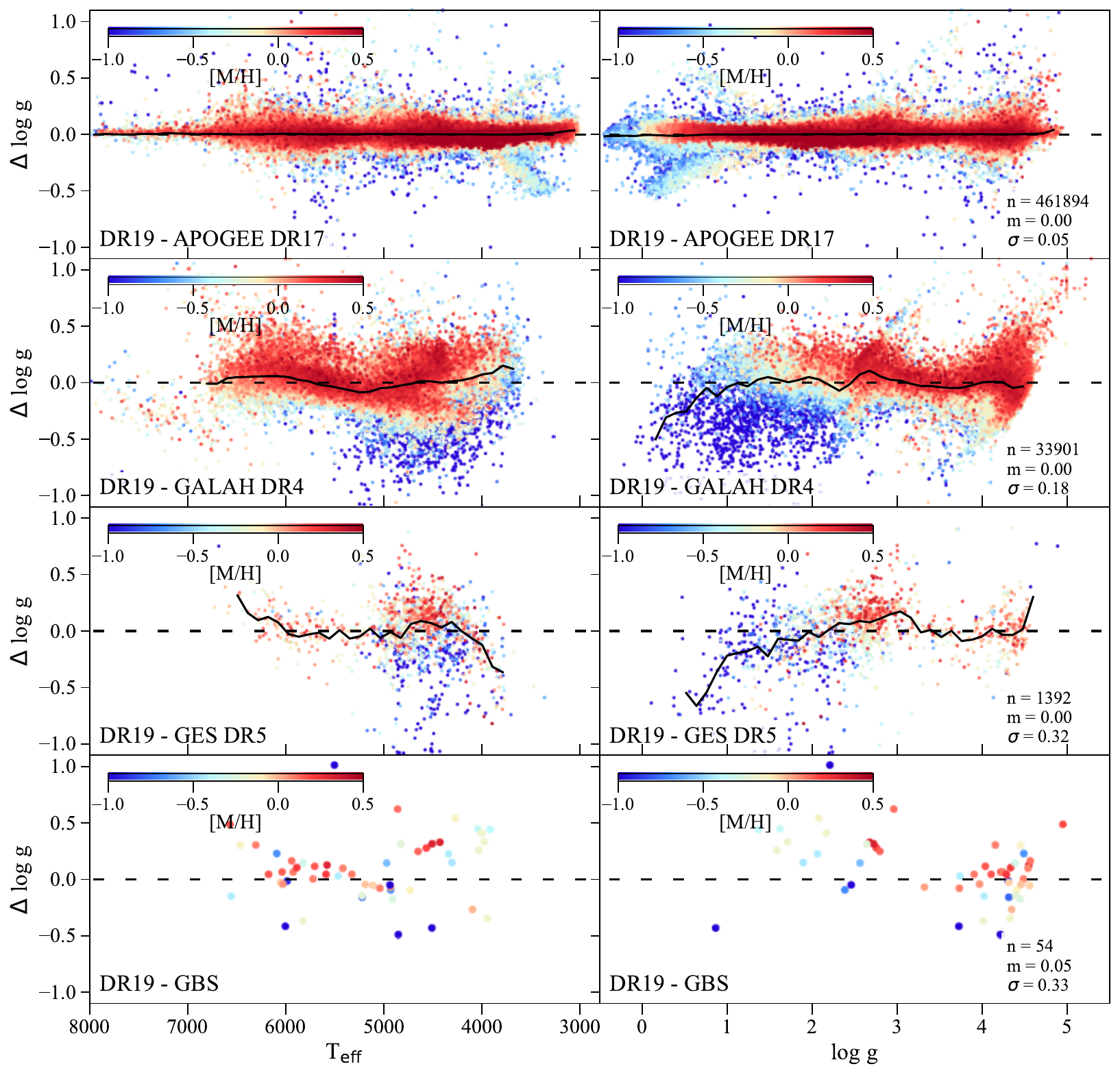}
\caption{Differences of DR19 ASPCAP raw surface gravities and APOGEE DR17 \citep{2022ApJS..259...35A}, GALAH DR4 \citep{2024arXiv240919858B}, Gaia-ESO DR5 \citep{2022A&A...666A.121R} and GBS \citep{2024A&A...682A.145S} values as a function of ASPCAP raw \teff \ color coded by raw metallicity. The solid black line shows the running median value.}
\label{fig:loggcomp}
\end{figure*}

The parameter range covered by the APOKASC3 and TESS samples is listed in Table~\ref{tab:validation-set}. The comparison of MWM \logg \ values with the asteroseismic ones can be seen in Figure~\ref{fig:loggcal}, where the top panel shows the RGB, the middle panel the RC stars, and the bottom panel indicates the comparison with the entire TESS sample regardless of evolutionary state. The classifications were taken from the APOKASC3 catalog. While the median discrepancy is only 0.09~dex for RGB stars, 0.18~dex for RC stars, and 0.14~dex in the full TESS sample. the difference between the spectroscopic and asteroseismic \logg \ also depends on \teff \ and \mh. The correlations were calibrated out in DR19 using the APOKASC3 sample as detailed by Casey et al. (2025, in preparation).

We caution users that abundances that are sensitive to surface gravity could have systematic offsets in the absolute abundance scale, because we use the spectroscopic surface gravities in the abundance analysis instead of the calibrated ones. RGB and RC stars had to be reliably identified in the main sample to properly assign the relevant calibration equation for each star. The method of how stars were separated into RGB and RC stars is explained by Casey et al. (2025, in preparation). Figure~\ref{fig:tefflogg} shows the Kiel-diagram using the raw \teff \ and \logg \ values. 

Since the uncertainty of seismic \logg \ is negligible compared to spectroscopic surface gravities, the overall uncertainty of spectroscopic surface gravities can be easily estimated from the standard deviation around the mean differences between the two datasets. These values are 0.07 and 0.09~dex for RGB and RC, respectively, which we consider to be excellent precision. Astra estimates uncertainties for each star, which is labeled raw\_e\_logg in the DR19 ASPCAP file. The median \logg \ uncertainty value reported in the entire sample is only 0.042~dex; thus, these values are underestimated compared to the \logg \ scatter observed in the APOKASC sample.

The APOKASC3 and TESS samples cover a narrow temperature range between 4000 and 5500~K, and metallicity range between $-$1.0 and 0.5~dex. A larger parameter space can be spanned with the comparison of \logg \ values from APOGEE DR17, GALAH DR4, Gaia-ESO DR5, and GBS stars, which are shown in Figure~\ref{fig:loggcomp}. The agreement with APOGEE DR17 is excellent across a large \teff \ and \logg \ range, and the median difference is essentially zero for the GALAH DR4 and GES DR5 samples, too. For the GBS sample, the median difference of 0.05~dex with a relatively large scatter of 0.33~dex based on 54 stars across all evolutionary phases. Considering the overall low values of our estimated uncertainties determined from the APOKASC sample, the large scatter seen here is probably the result of the low number of GBS stars in the sample.

\subsubsection{M Dwarfs}
\label{mdwarflogg}

The APOKASC3 catalog does not contain any main-sequence stars below 4500~K, so it is not possible to analyze the quality of the MWM \logg \ values in this parameter region. Thus, Dartmouth isochrones sampled at 12 Gyr old \citep{2008ApJS..178...89D} act as a reference at metallicities spanning $-$1.5 and 0.5~dex shown in the Kiel diagram in Figure~\ref{fig:tefflogg}. The ASPCAP surface gravities show a near-constant value below 5000~K, while the isochrones exhibit the expected increase of \logg \ with decreasing \teff \ for K and M dwarfs. This results in an increasing discrepancy between the ASPCAP and isochrone \logg \ starting to be significant below 4500~K reaching almost 1~dex between 3000 and 3500~K, which may also affect metallicity and individual abundances of dwarfs below 4500~K as detailed in Section~\ref{mdwarfmet} and Section~\ref{sec:abundances} (see also \citet{2023ApJ...951...90W}). 

The difference between the surface gravities derived by \citet{2022ApJ...927..123S} and ASPCAP DR19 is shown in the middle panel of Figure~\ref{mdwarfs}, and exhibits a very similar systematic offset up to about 1~dex. Thus, raw values of \logg \ above 4.0 and \teff $<$ 4500~K should not be trusted, but surface gravities were calibrated to isochrones as detailed by Casey et al. (2025, in preparation). The exact cause of the incorrect surface gravities is currently not fully understood, but most likely the result of incorrect physical assumptions made during the calculations of the synthetic spectral libraries.



\subsection{Metallicity}
\label{sec:metallicity}

\begin{figure}                          
\centering
\includegraphics[width=3.42in,angle=0]{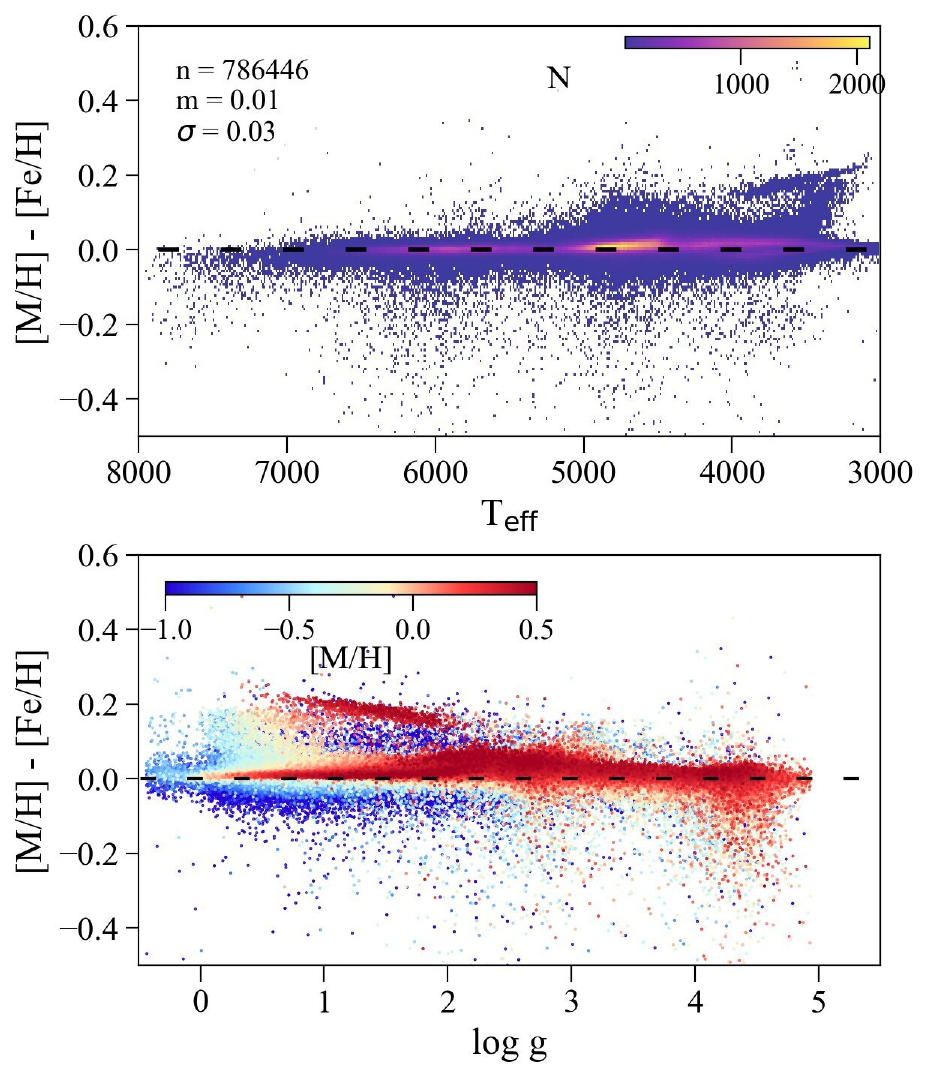}
\caption{Top panel: the 2D histogram of the difference between \mh \ and \feh \ as a function of raw \teff. Bottom panel: the difference between \mh \ and \feh \ as a function of raw \logg \ color coded by \mh. Only stars with SNR$>$50 and no bad flags are shown.}
\label{fig:metfehcomp}
\end{figure}

\begin{figure*}                          
\centering
\includegraphics[width=7.1in,angle=0]{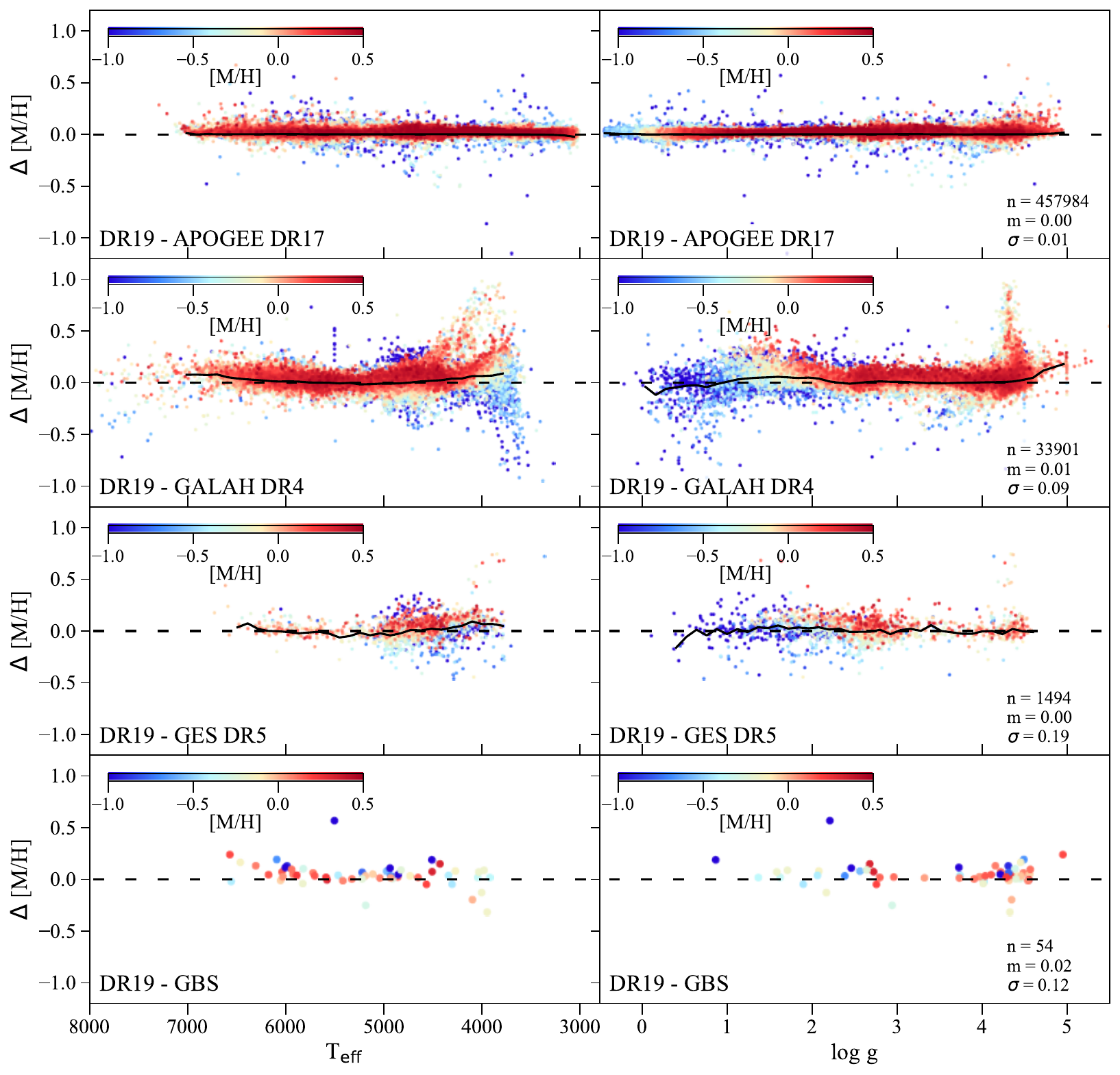}
\caption{The difference between MWM \mh \ and literature \feh \ as a function of ASPCAP raw \teff \ and \logg \ from APOGEE DR17 (top panels), GALAH DR4 (second row from the top), Gaia-ESO DR5 (second row from the bottom) for stars with SNR$>$50 from both surveys, and no BAD flags from MWM. The bottom panels show the comparison with the GBS sample. The solid black line shows the running median value.}
\label{fig:metcomp}
\end{figure*}

The abundance of each individual element X heavier than helium is defined as:

\begin{equation}
\mathrm{[X/H]} = \log_{10}(n_X/n_H)_{\rm star} - \log_{10}(n_X/n_H)_\odot 
\end{equation}

\noindent where $n_X$ and $n_H$ are, respectively, the number of atoms of element X and hydrogen, per unit volume in the stellar photosphere. Following the original APOGEE definition of metallicity \citep{2013AJ....146..133M}, we define [M/H] as an overall scaling of metal abundances with a solar abundance pattern, and [X/M] as the deviation of element X from the solar abundance pattern:

\begin{equation}
\mathrm{[X/M]} = \mathrm{[X/H]} - \mathrm{[M/H]}.
\label{eqn:mh}
\end{equation}
For FGKM stars, \mh \ is generally very close to the \feh \ that is often used in the literature instead of \mh. \feh \ is derived in the second stage of ASPCAP in wavelength windows that are sensitive to only lines of Fe; it is stored in raw\_fe\_h in the DR19 ASPCAP file. The similarity of the two is because iron-group elements tend to closely vary together in general, and the number of Fe lines in the $H$ band is quite large.

Figure~\ref{fig:metfehcomp} illustrates the relation between the two metallicity scales for stars with SNR $>$ 50 and no bad flags. Based on nearly 800,000 stars, the median difference is 0.01~dex with a scatter of only 0.03~dex and thus, for most of the targets, the two metallicities are within the expected uncertainties. Interestingly, a small group of nearly 1500 stars can be identified that deviate from the main survey sample and have differences between 0.13 and 0.22~dex. They can be seen in the bottom panel of Figure~\ref{fig:metfehcomp}. This group is unique in that their \teff \ lies between 3300 and 4000~K, \logg \ between 0.5 and 2, and are all metal rich with \mh \ $>$ 0~dex. The exact reason for this discrepancy is currently unknown.

The main reason for the differences between \mh \ and \feh \ is that, unlike \feh, \mh \ can be influenced by other elements with strong absorption lines in a given stellar spectrum, such as carbon or $\alpha$ elements. In DR19, both values are reported to allow the user to adopt \mh \ or \feh \ as the metallicity value in their studies. However, we will continue using \mh \ for our analyses because that is the value that was fixed in the global fit, and then used to derive the individual abundances. 

We compare our overall metallicity scale, \mh, with the metallicities obtained for stars observed in common with APOGEE DR17 \citep{2022ApJS..259...35A}, GALAH DR4 \citep{2024arXiv240919858B} and Gaia-ESO DR5 \citep{2022A&A...666A.121R}. The overall parameter ranges and number of stars for this comparison sample can be found in Table~\ref{tab:validation-set}. The agreement is excellent with APOGEE DR17 (the median difference is 0.0, the scatter is 0.01~dex), and no systematic differences can be seen as a function of \teff \ and \logg \ (see top panels of Figure~\ref{fig:metcomp}).

We note that in GALAH, the metallicity parameter, \feh\, represents a pseudo-iron abundance from the best-fit atmospheric model, estimated from Sc, Ti, and Fe lines \citep{2021MNRAS.506..150B}. In the Gaia-ESO Survey, on the other hand, the metallicity \feh \ corresponds to a combination of estimates that come from measurements obtained using various methods that span from equivalent width determination of Fe lines to global fitting of synthetic spectra with SME and FERRE \citep{2024arXiv240206076W}. While the median difference between GALAH is 0.01, and Gaia-ESO is 0.0~dex, the agreement between MWM and GALAH strongly depends on temperature and \logg, as can be seen in Figure~\ref{fig:metcomp}. The disagreement increases with decreasing temperature below 4500~K for both metal poor and metal rich RGB and RC stars (but not MS stars), where strong molecular bands of TiO start to form at optical wavelengths, which makes the continuum-normalization difficult \citep{2012A&A...547A.108L, 2017A&A...601A..38J}.

The agreement between MWM and Gaia-ESO is much better, as there are no significant correlations between their two metallicity scales with \teff \ or \logg \ (middle panels of Figure~\ref{fig:metcomp}). The lack of any correlations between the infrared and optical metallicities with \teff \ and \logg\ is reinforced by the differences seen in the GBS sample (bottom panels of Figure~\ref{fig:metcomp}), where the agreement is again excellent; the median difference is only 0.02~dex. It is worth noting that the number of stars in the MWM-GES and GBS samples is much lower than in the MWM-GALAH catalog; thus, exploring detailed correlations in those two samples is limited.

To estimate the overall uncertainty in metallicity, we selected stars from open clusters part of the OCCAM survey's APOGEE DR17 version \citep{2018AJ....156..142D, 2020AJ....159..199D, 2022AJ....164...85M} and globular clusters from \citet{2020MNRAS.492.1641M} using the membership selection of the literature. Globular clusters with known iron spread were removed from our list. To select a clean sample, stars with \snr \ $>$ 50, \logg \ $<$ 3.8, 4000~K $<$ \teff $<$ 6000~K, and no bad flags were kept in the sample. Only clusters with at least five members were selected. The temperature cut was introduced to limit the parameter space to the region where it is the most precise to measure abundances in the $H$ band \citep{2020AJ....160..120J}. Dwarf stars are not used here, because MWM has not observed enough main-sequence stars so far to reliably estimate the scatter or derive any temperature-dependent calibrations. Restricting the sample to giant stars further allowed us to avoid the turnoff, where diffusion takes its largest effect \citep{2008A&A...490..777L, 2018ApJ...857...14S}. The final sample contained 358 stars in 12 open clusters, and 1344 stars in 16 globular clusters. The general parameter range that these open and globular cluster samples cover is listed in Table~\ref{tab:validation-set}. The average scatter in the 12 open clusters is only 0.033~dex, while in the globular clusters, this value reaches 0.092~dex.

\begin{figure}                          
\centering
\includegraphics[width=3.42in,angle=0]{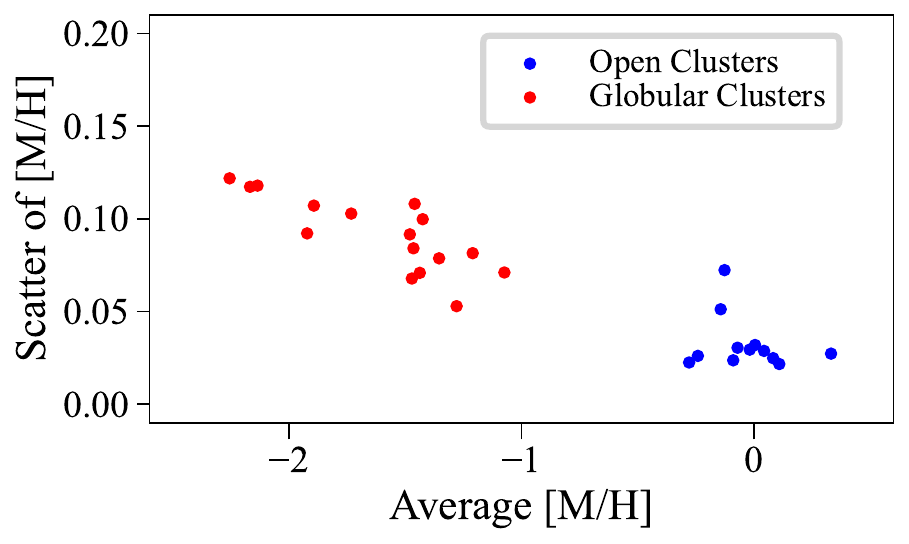}
\caption{Scatter of metallicity in 12 open and 16 globular clusters. A slight anticorrelation with average \mh \ can be seen as lines in $H$ band get weaker with decreasing metallicity.}
\label{fig:metsca}
\end{figure}

Some of the scatter observed in open clusters may have astrophysical origins. Recent high-precision abundance analysis of open clusters can reach a star-to-star scatter of the order of 0.01 dex \citep[e.g.][]{2021A&A...654A.151C, 2024ApJ...975...89S}. Below that value, star-to-star differences can also be attributed to real inhomogeneities due to, e.g. mixing and gravitational settling processes or planet engulfment \citep{2016MNRAS.457.3934L, 2018ApJ...857...14S}, and possibly due to inhomogeneous chemical composition of the molecular cloud that gave the cluster birth \citep{2024A&A...682L..23S}. Clusters members are more homogeneous than field stars, but the exact level of homogeneity still remains to be quantified. For these reasons, the uncertainties shown in Figure~\ref{fig:metsca} serve only as an upper limit of our estimated precision.

\vspace{0.5cm}

\subsubsection{M Dwarfs}
\label{mdwarfmet}

The large offset of the surface gravities of stars with \teff \ $<$ 4500~K and \logg \ $>$ 4.0 compared to isochrones (see Section~\ref{secgrav}) may result in incorrect metallicities for at least some part of the M dwarf sample. This can be seen in Figure~\ref{fig:tefflogg}, where some of the red dwarf stars appear to have metallicities lower than $-$0.5~dex below 3800~K. Qiu et al. (in preparation) compared the DR19 ASPCAP metallicities with SLAM \citep{2020ApJS..246....9Z} metallicities for M dwarfs with both APOGEE and BOSS spectra. They also found that ASPCAP metallicities were too low for the coolest stars. This is also reinforced by the difference between metallicities derived independently from ASPCAP by \citet{2022ApJ...927..123S} and ASPCAP DR19 values shown in the bottom panel of Figure~\ref{mdwarfs}. The median value of the discrepancy between the two metallicity scales is 0.18~dex, and the offset is independent of \teff \ and \mh. 

Thus, the ASPCAP DR19 metallicity of at least some of these stars is incorrect, which may be due to inaccuracies in the synthetic spectral libraries that are used to fit the observed spectra. While the effective temperature of solar-metallicity M dwarfs seems to be reliable (see Section~\ref{mdwarfteff}), the offset in surface gravity compared to isochrones and \citet{2022ApJ...927..123S} (Section~\ref{mdwarflogg}) combined with the metallicity offset presented here may also affect the abundances of most M dwarfs, so we caution users to carefully check abundances of stars with \teff \ $<$ 4500~K and \logg \ $>$ 4.0 before using them for any science application. Although these issues may be present in DR19, \citet{2022ApJ...927..123S} found that abundances of C, O, Na, Mg, Al, Si, K, Ca, Ti, V, Cr, Mn, Fe, and Ni can be reliably measured from APOGEE spectra of M dwarfs with careful manual analysis.


\section{Chemical Abundances in DR19}
\label{sec:abundances}

\begin{figure*}                          
\centering
\includegraphics[width=7.1in,angle=0]{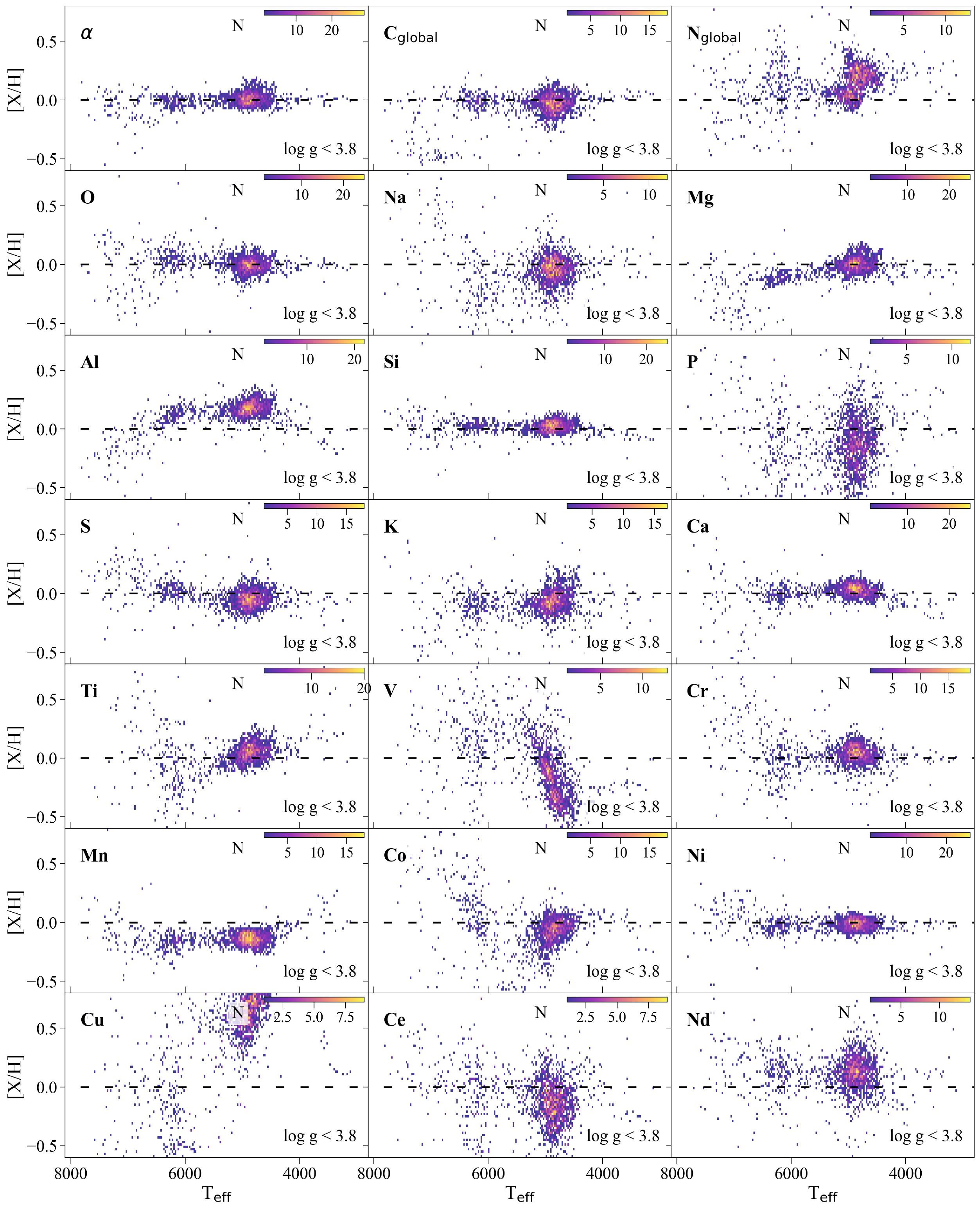}
\caption{Raw abundances of giants in the solar neighborhood solar-metallicity sample as a function of \teff. Stars with \snr \ $>$ 50, no BAD flags, and $-$0.05 $<$ \mh \ $<$ 0.05 are plotted only. The zero-point offset for giants stars were determined from the 4000$-$6000~K range. Abundances of C$_{\rm global}$ and N$_{\rm global}$ are from the global fit.}
\label{solarneigh}
\end{figure*}

\begin{figure*}                          
\centering
\includegraphics[width=7.1in,angle=0]{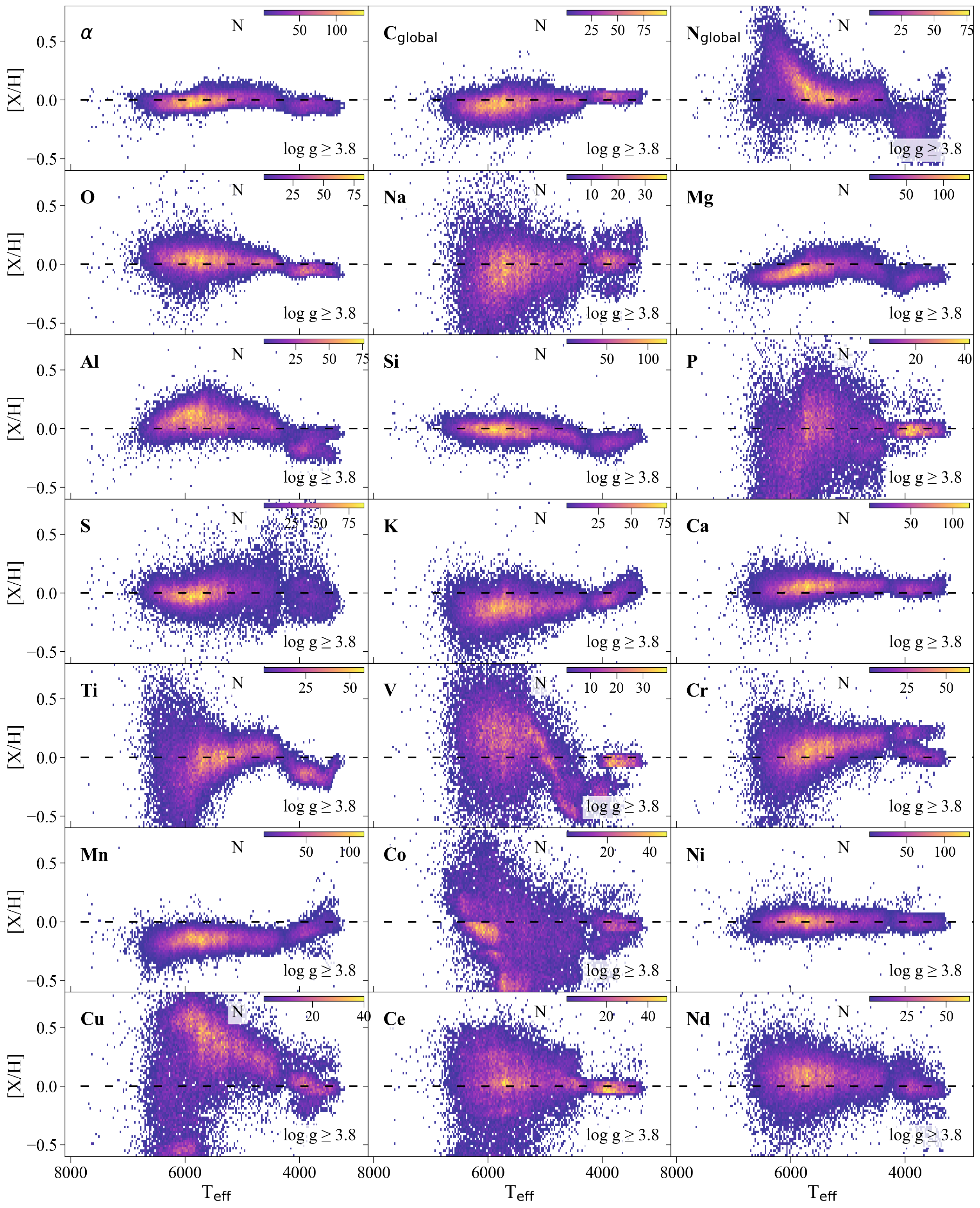}
\caption{Same as in Figure~\ref{solarneigh} but for dwarfs.}
\label{solarneigh2}
\end{figure*}

\begin{figure*}                         
\centering
\includegraphics[width=7.1in,angle=0]{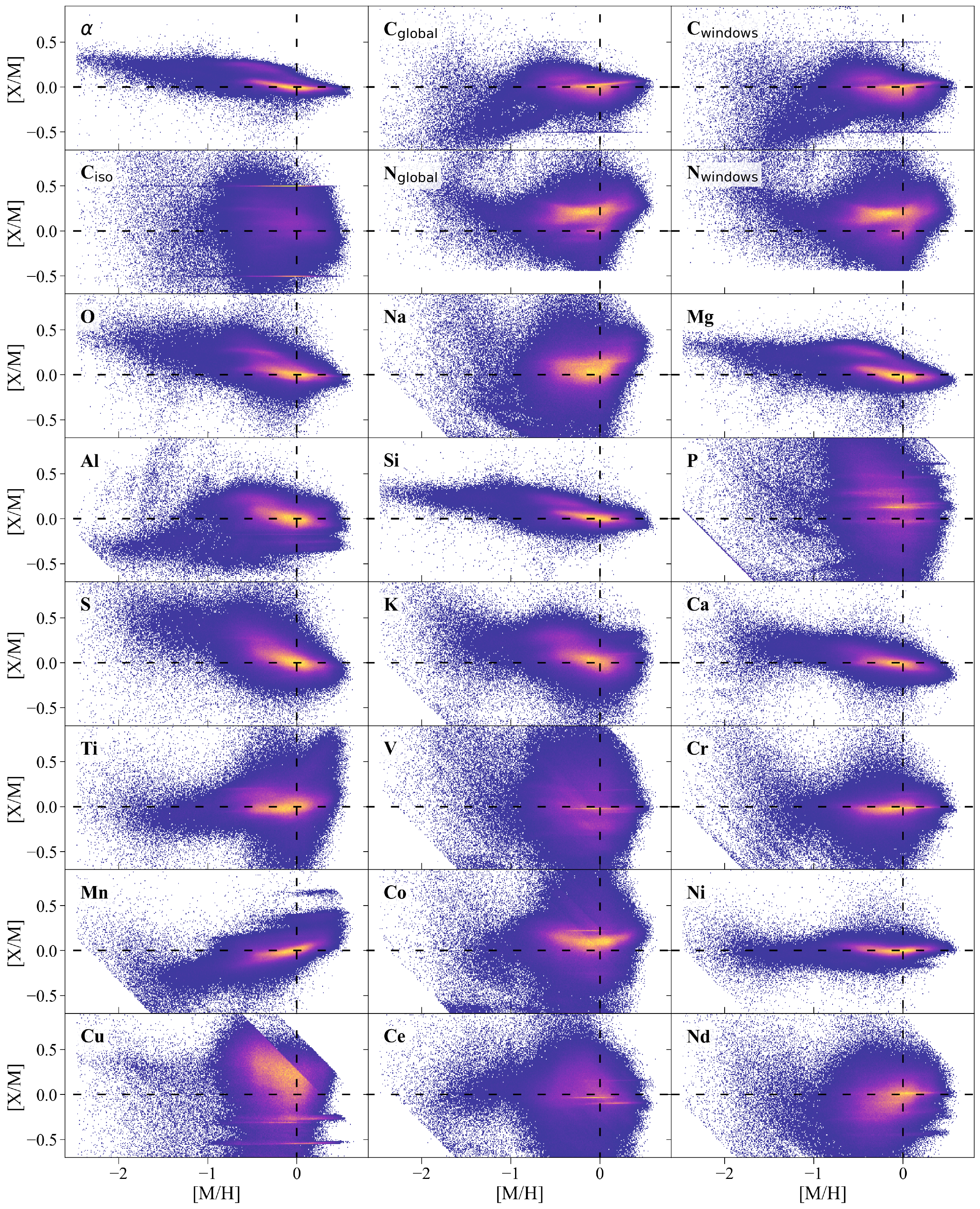}
\caption{Corrected abundances published in DR19 as a function of \mh \ of stars with \snr \ $>$ 50, no BAD flags, and e$\_$x$\_$h $<$ 0.2. Features seen in the calibrated abundances are discussed in the individual abundance sections. Stripes seen in the low-density regions of some abundances (Na, P, K, V, Cr, Co, Cu, Ce, Nd) indicate noding problems at low abundance values. All abundances are plotted, except those that are within 0.03 dex of the grid edge. Abundances of C$_{\rm global}$ and N$_{\rm global}$ are from the global fit.}
\label{abunmap}
\end{figure*}

Individual abundances of elements are derived in the second phase of ASPCAP by fixing the main atmospheric parameters and using wavelength windows sensitive to the particular element. To be consistent with the parameters that were adopted from the best-fit spectrum, we used the raw \teff\ and \logg\ values to derive abundances, because these give the best overall fit to the stellar spectrum. They also provide more consistent results for molecular blends like CO, CN, and OH that are present in the spectra, which is especially important when determining the abundances from the windows.

The list of elements whose abundances are derived in the second phase include carbon and nitrogen, but not $\alpha$, as the abundance of each $\alpha$ element is derived separately. Overall, \xh \ values of 24 different abundances were derived for 21 different elements (including two metallicity definitions) in DR19, because some elements have multiple entries. The raw abundance values are stored under the raw$\_$x$\_$h nametags, where x marks the individual element as listed in the specific abundance subsection here. Elements except carbon and nitrogen also have zero-point offsets applied; these calibrated values are stored under the x$\_$h nametag. 

In this section, we first describe our strategy for improving the accuracy of our abundances by using the solar-metallicity solar neighborhood stars from Table~\ref{tab:validation-set}. We then evaluate the dependency of the abundances on temperature, and describe how our estimated precision was calculated using various samples, which are listed in Table~\ref{tab:validation-set}. This analysis allows us to provide a quality assessment of the abundances, as well as to evaluate the parameter space where a particular abundance might have biases. Detailed discussions can be found in the dedicated subsections of individual abundances in Section~\ref{elements}.

\subsection{Accuracy of Abundances}
\label{accuabusec}

\begin{deluxetable}{lrr}
\tabletypesize{\scriptsize}
\tablecaption{Abundance Zero-Point Offsets Applied in DR19.}
\tablewidth{0pt}
\tablehead{
\colhead{Element} & \colhead{Giants} & 
\colhead{Dwarfs}  \\ 
\colhead{} & \colhead{\logg $<$ 3.8} & 
\colhead{\logg \ $\geq$ 3.8}}
\startdata
\am   &   0.0088     &    0.0024     \\
\oh   &  $-$0.0039   &   0.0236       \\
\nah  &  $-$0.0507   &   $-$0.0497    \\
\mgh  &  $-$0.0005   &   $-$0.0305    \\
\alh  &   0.1751     &    0.0745      \\
\sih  &   0.0234     &   $-$0.0301    \\
\ph   &  $-$0.1081   &   $-$0.0121    \\
\sh   &  $-$0.0446   &    0.0053         \\
\kh   &  $-$0.0662   &   $-$0.1099       \\
\cah  &   0.0323     &    0.0631      \\
\tih  &   0.0470     &   $-$0.0099       \\
\vh   &  $-$0.1383   &    0.0233      \\
\crh  &   0.0317     &    0.0694      \\
\mnh  &  $-$0.1338   &   $-$0.1444    \\
\coh  &  $-$0.0690   &   $-$0.1047    \\
\nih  &  $-$0.0127   &    0.0101      \\
\cuh  &   0.3230     &    0.2138      \\
\ceh  &  $-$0.1186   &    0.0331         \\
\ndh  &   0.1185     &    0.0705      \\
\enddata 
\label{zeropoint}
\end{deluxetable}

To assess the accuracy of the DR19 abundances, we compare the raw abundances with solar-metallicity stars from the solar neighborhood sample (SNSM; see Table~\ref{tab:validation-set}). This is the same procedure followed by the APOGEE survey in DR17  \citep{2022ApJS..259...35A}. We expect discrepancies in our results because obtaining objectively accurate abundance measurements is fundamentally difficult. This is due to their dependence on assumptions made in the solar reference abundance scale, the model atmosphere calculations, synthesis, astrophysical data used in the line list, small inaccuracies in the continuum-normalization, and other factors, as extensively reviewed by \cite{2019ARA&A..57..571J}.

The choice of using the SNSM sample as "ground truth" against which to evaluate the accuracy of abundances was motivated by the expectation that these stars have very similar \xm \ values with respect to the Sun. This follows the assumption that low-mass stars retain their original chemical composition in their spectra as they age \citep{2002ARA&A..40..487F}, and so the abundance distribution of the solar neighborhood should not correlate with \teff\ or \logg. Literature results also suggest that the mean \xm \ values, thus also \xh, in solar-metallicity solar neighborhood stars are close to the solar values \citep{2003MNRAS.340..304R, 2012A&A...545A..32A, 2014A&A...562A..71B}, regardless of their stellar parameters. Abundances of C and N, however, can be modified by stars when they ascend the RGB and mix their interior material with the photosphere \citep[][and references therein]{2019A&A...621A..24L}. These abundances are therefore not included in this discussion nor are they calibrated.

To select a sample of stars in the solar vicinity that are assumed to form a single population, \mh \ was restricted to be within $\pm$ \ 0.05~dex, the Gaia DR3 \citep{2023A&A...674A...1G, 2016A&A...595A...1G} parallax to be larger than 0.02 arcsec (distance smaller than 500 pc), \snr \ to be larger than 50 and no BAD flags were allowed. The overall properties of the SNSM sample are listed in Table~\ref{tab:validation-set}. 

The applied calibration for DR19 is a simple zero-point offset, $\Delta$, that is determined by forcing the SNSM sample to have a mean \xh \ = 0. Since line strength often depends on surface gravity, the calibration was done separately for giants (\logg$ \ <$ 3.8) and dwarfs (\logg$ \ \ge$ 3.8). Stars that have  \teff \ between 4000~K and 6000~K have been selected from the solar-metallicity solar neighborhood sample, because this is the temperature region where most absorption lines correlate with the abundance of elements \citep{2005oasp.book.....G}, and outside this region, large offsets and inaccurate abundance measurements are more likely. For more details, we refer the reader to the discussion of individual elements in the subsections below.

These zero-point offsets are listed in Table~\ref{zeropoint}. These offsets are, in most cases, below 0.1 dex, although there are some exceptions, like Al, Mn, and Cu. The calibrated \xh \ values can be calculated by subtracting the offsets $\Delta$ values found in this table from the raw \xh \ values. The raw \xh \ abundances in the solar neighborhood sample are shown in Figure~\ref{solarneigh} and Figure~\ref{solarneigh2} for giants and dwarfs separately. We remind the reader that C and N are not calibrated because we do not expect these abundances to scale to solar in giants due to mixing processes. Thus, C and N are not shown in these figures. The solar neighborhood sample is dominated by dwarf stars covering a large temperature range. There are relatively few giants within 500 pc below 4000~K, which makes it difficult to assess the accuracy of abundances in the atmosphere of cool giants. 

The calibrated abundances have a distribution centered around solar values by construction, but dependencies with temperature are present in most elements, in case of main-sequence stars. Since no obvious patterns can be quantified from these trends, we follow previous data releases of APOGEE and do not apply further calibrations to the abundances of dwarf stars. Giant stars, shown in Figure~\ref{solarneigh}, although less numerous than dwarf stars, show less prominent trends with temperature. These various distributions are further discussed when each individual element is analyzed in the sections below.

The distribution of zero-point calibrated abundances for the entire DR19 sample as a function of \mh \ can be seen in Figure~\ref{abunmap}. Here, we only plot stars with \snr \ $>$ 50, no individual element BAD flags (x$\_$h$\_$flags = 0), and small reported internal Astra uncertainty (e$\_$x$\_$h $<$ 0.2). In general, the distribution of abundances follows previous spectroscopic surveys \citep{2020AJ....160..120J, 2021MNRAS.506..150B, 2023A&A...674A...1G, 2016A&A...595A...1G} and our expectations of chemical evolution in the Milky Way. The $\alpha$-capture elements such as O, Mg, Si, and Ca have an overall enhancement of about 0.2$-$0.4 dex for metal poor stars, and the distribution presents its knee at about [M/H] of $-$1. Other iron-peak elements such as Cr and Ni show the expected flat trends as a function of metallicity, while others like Mn show an increasing trend. There are elements with high dispersion, such as Nd, P, K, Cu, Ce and Na. Before further discussing the reason for these high dispersions, the temperature dependence and precision of all abundances have to be first assessed.

\subsection{Temperature Dependency of the Corrected Abundances}
\label{abuteffsec}

\begin{deluxetable}{lrrrrr}
\tabletypesize{\scriptsize}
\tablecaption{Abundance Correction Coefficients as a Function of \teff.}
\tablewidth{0pt}
\tablehead{
\colhead{Element} & \colhead{$a$} & 
\colhead{$b$} & \colhead{Offset} & \colhead{Offset} & 
\colhead{$\delta$} \\ 
\colhead{} & \colhead{} & 
\colhead{} & \colhead{$>$ 6000~K} & \colhead{$<$ 3500~K} & \colhead{}
}
\startdata
\am       & $-$2.2918e-5  &   0.0861  &   $-$0.0514  & 0.0059     & $-$0.0458  \\
\om       & $-$4.0909e-5  &   0.1651  &   $-$0.0804  & 0.0219     & $-$0.0818  \\
\nam      & $-$8.2173e-5  &   0.4586  &   $-$0.0344  & 0.1710      & $-$0.1643  \\
\mgm      & $-$4.3932e-5  &   0.1733  &   $-$0.0903  & 0.0195     & $-$0.0879  \\
\alm      &    3.0850e-5  &  $-$0.1734  & 0.0117   &   $-$0.0654  & 0.0617   \\
\simm     &    1.1688e-5  &  $-$0.0594  & 0.0107   &   $-$0.0185  & 0.0234  \\
\pmm      & $-$2.1574e-4  &   1.0477  &   $-$0.2467  & 0.2926     & $-$0.4315  \\
\sm       & $-$1.0886e-6  &   0.0214  &   0.0149   &   0.0176     & $-$0.0022  \\
\km       & $-$7.0765e-5  &   0.3220  &   $-$0.1026  & 0.0743     & $-$0.1415 \\
\cam      &    5.4495e-5  &  $-$0.2832  & 0.0438   &   $-$0.0925  & 0.109   \\
\tim      & $-$1.1466e-4  &   0.5137  &   $-$0.1743  & 0.1124     & $-$0.2293  \\
\crm      & $-$6.4991e-6  &   0.0099  &   $-$0.0291  & $-$0.0128  & $-$0.013   \\
\mnm      & $-$1.0168e-4  &   0.4999  &   $-$0.1102  & 0.1440      & $-$0.2034  \\
\com      & $-$1.8123e-4  &   0.8724  &   $-$0.215   & 0.2381     & $-$0.3625 \\
\nim      & $-$2.3203e-5  &   0.0806  &   $-$0.0586  & $-$0.0006  & $-$0.0464  \\
\cem      &    1.3833e-4  &  $-$0.5431  & 0.2869   &   $-$0.0589  & 0.2767   \\
\ndm      &    6.4019e-5  &  $-$0.2996  & 0.0845   &   $-$0.0755  & 0.128 \\
\enddata
\label{corrcoeff}
\end{deluxetable}

\begin{figure*}                          
\centering
\includegraphics[width=7.1in,angle=0]{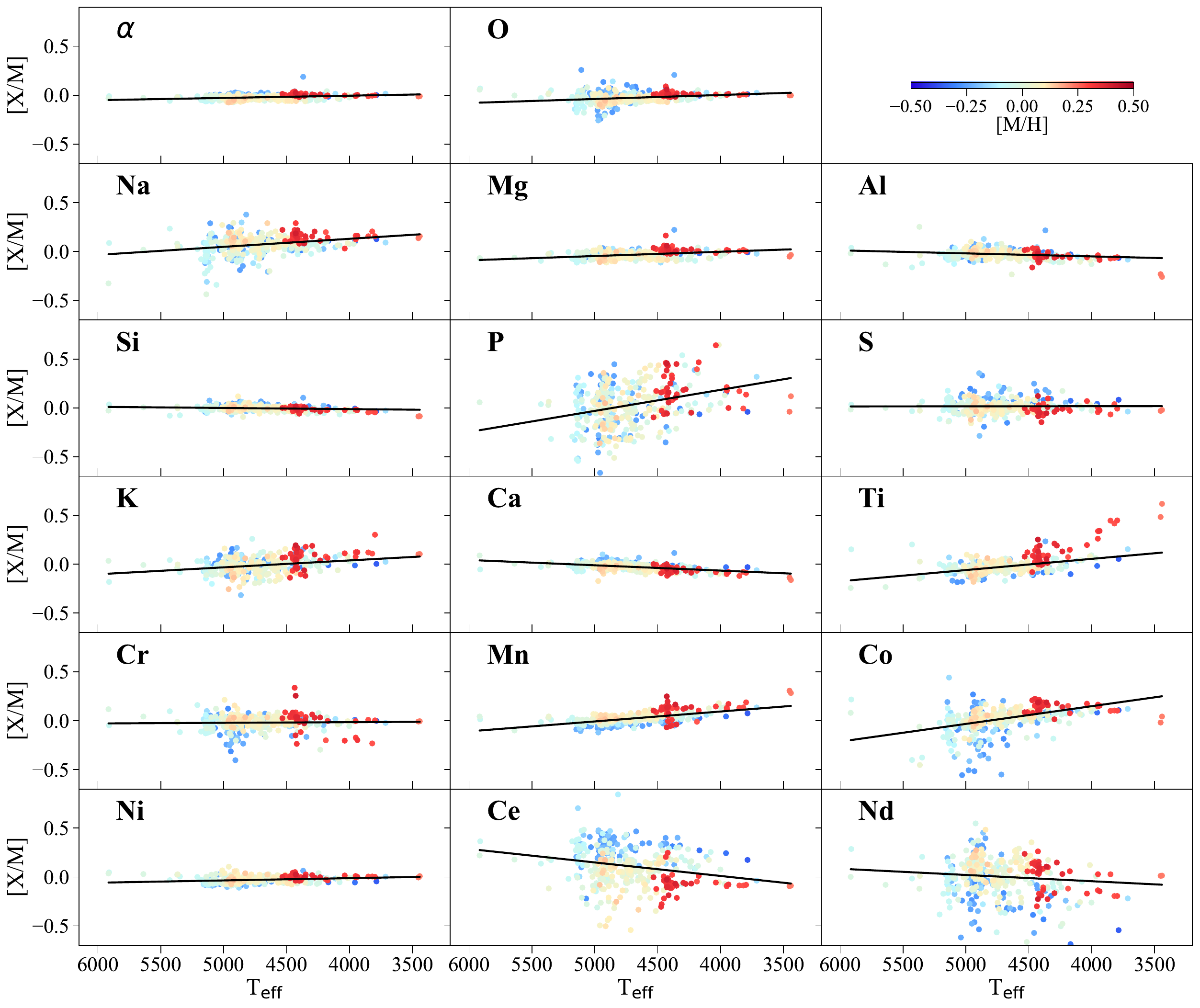}
\caption{\xm \ abundances as a function of \teff \ using 358 giant stars in 12 open clusters from the OCCAM survey. The solid line shows the derived calibration equation. Stars with \snr \ $>$ 50, \logg \ $<$ 3.8, \teff $<$ 6000~K, and no BAD flags were kept in the sample.}
\label{abuteff}
\end{figure*}

Any possible correlation between abundances and the main atmospheric parameters can be checked using open clusters, in which all stars are formed in a very narrow timespan out of the same cloud of gas and dust; thus, they have very similar abundances for all elements. As such, we expect that abundances as a function of temperature exhibit constant values. Exceptions to this assumption are C and N, because of arguments explained in Section~\ref{accuabusec}. Following this logic, we expect that \xm \ values in open clusters should show a constant value as a function of \teff \ and \logg, regardless of their metallicity. However, recently \citet{2018ApJ...857...14S} showed that diffusion, which is expected in main-sequence atmospheres, can affect the abundance of multiple elements near the turnoff in M67 in APOGEE data. We need to take this effect into account when deriving temperature-based dependencies.

Such temperature-based corrections were applied in DR14 \citep{2018AJ....156..125H}, but not in DR16 \citep{2020AJ....160..120J} and DR17 \citep{2022ApJS..259...35A}. The groups that needed to restrict the range of their sample in \logg \ or \teff \ have applied their own corrections \citep{2021ApJ...909...77G, 2022ApJS..260...32W, 2024ApJ...970..180S}, which resulted in corrections that differed from each other. In DR19, we aim to present a uniform set of temperature-based corrections.

To quantify possible dependencies of the abundances with \teff, we utilize the open cluster sample (see Table~\ref{tab:validation-set}), which was also used to validate the metallicity in Section~\ref{sec:metallicity}. The \teff\ range of the open cluster sample also ensures that the most-precise abundances are used, because observed stars in the selected open clusters have temperatures between 3500 and 6000~K, a range in which most of the absorption lines lie in the optimal part of the curve of growth \citep{2005oasp.book.....G}. At the end, all 351 members from the 12 clusters were combined, and their abundance distribution as a function of \teff \ is plotted in Figure~\ref{abuteff}. Then, the individual abundances of these open cluster stars are used to perform a linear regression fit with \teff\ to determine the coefficients $a$ and $b$ as follow: 

\begin{equation}
\Delta \mathrm{[X/M]} = a \cdot \mathrm{T_{\rm eff}} + b
\label{teffabueq}
\end{equation}

The values of the $a$ and $b$ coefficients are listed in Table~\ref{corrcoeff}, and the lines are plotted in Figure~\ref{abuteff}. In principle, one could consider more accurate abundances by calculating new \xm \ values by \xm$_{\rm calib.} - \Delta$ \xm. This correction  only applies for stars with \logg \ $<$ 3.8, and 3500~K $<$ \teff \ $<$ 6000~K. Outside this extended temperature region, we recommend to use a constant offset listed in Table~\ref{corrcoeff} that are calculated using Equation~\ref{teffabueq} at 6000~K and 3500~K, respectively. The \teff-dependent coefficients are provided for the following elements: $\alpha$, O, Na, Mg, Al, Si, P, S, K, Ca, Ti, Cr, Mn, Co, Ni, Ce, and Nd. We do not provide temperature based corrections for the abundances of P, V, and Cu, because the high scatter observed in open clusters does not make it possible to reliably establish a correlation with \teff. It is important to  stress that calibrations based on these coefficients are not included in the DR19 ASPCAP file. The source of the correlation of abundances with \teff \ is not fully understood, but among the possible causes are the usage of uncalibrated surface gravities in the abundance determination, the lack of NLTE and/or 3D corrections that may depend on \teff, or the fact that the line strength of different absorption lines of the same element are affected by \teff \ differently.

\subsection{Precision of Abundances}
\label{precabu}

In this paper, the precision of the abundances is evaluated using three different samples. These are the solar neighborhood, open clusters, and wide binaries, whose general properties are listed in Table~\ref{tab:validation-set}. The estimated uncertainty corresponds to the mean scatter of each abundance, which is independent from the uncertainty that is calculated from the repeated observations. The solar neighborhood sample consists of the same solar-metallicity stars used for the zero-point offset calibration (see Section~\ref{accuabusec}). Since the metallicity of these stars spans $\pm$ 0.05~dex, we needed to subtract 0.035 in quadrature when calculating the scatter of abundances in this sample. Thus, this criterion estimates the uncertainty of \xm, not \xh. 

\begin{figure*}                          
\centering
\includegraphics[width=7.1in,angle=0]{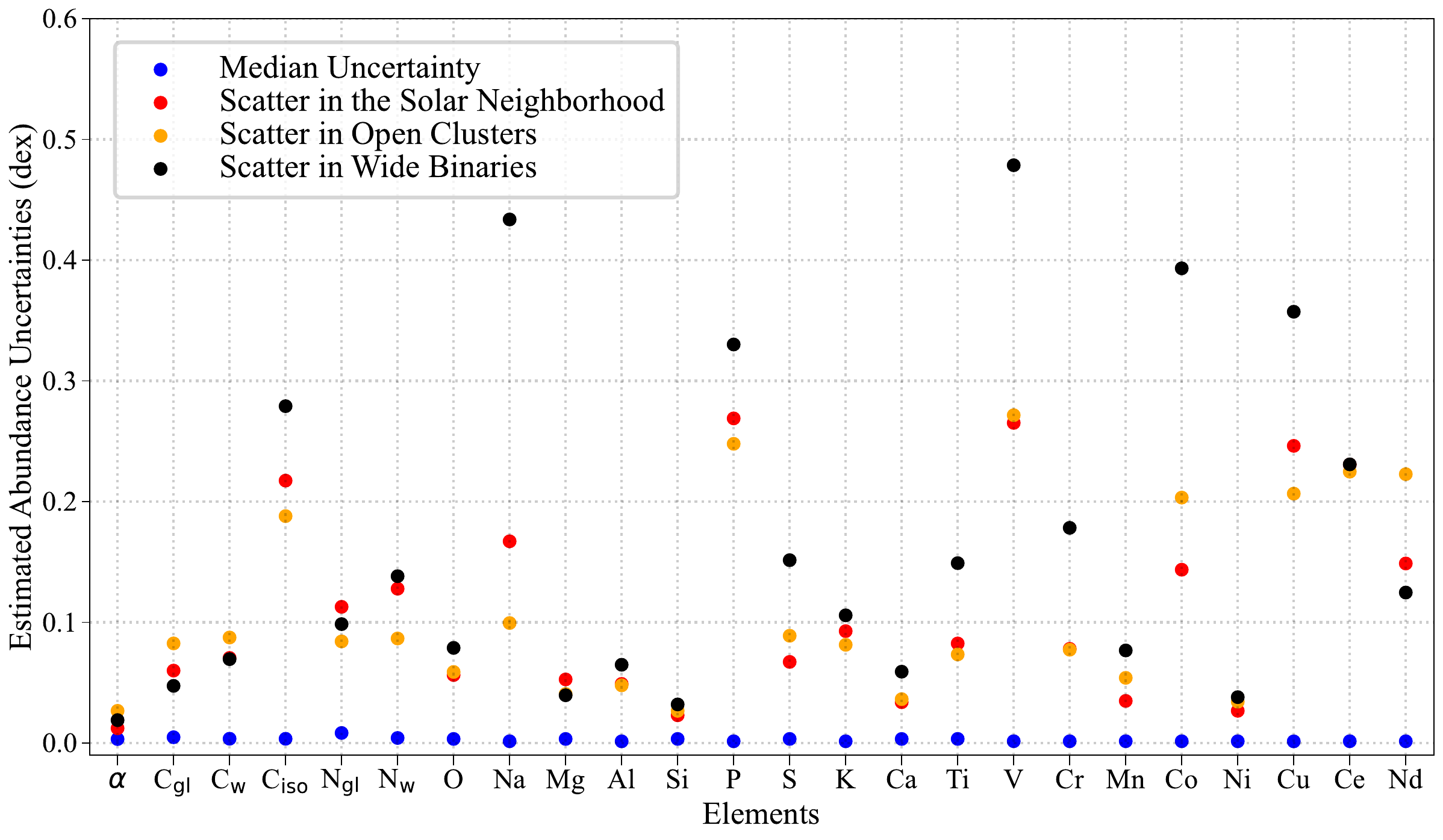}
\caption{The median DR19 uncertainties (blue dots) for each element compared to scatter measured in the solar neighboorhod sample (red dots), in 12 open clusters (yellow dots), and 348 wide binaries (black dots). The range of \teff \ is between 4000 and 6000~K, \snr \ $>$ 50.}
\label{abuerrors}
\end{figure*}

The open cluster sample considers only giant stars; however, it cannot be applied to C and N due to deep mixing, and the turnoff must be avoided because of diffusion. We use the same set of stars that were selected in Section~\ref{abuteffsec}, but further restrict the temperature range to only contain stars hotter than 4000~K, but cooler than 6000~K for the same reasons the SNSM sample was limited to this temperature range. Since each open cluster has a slightly different metallicity, the scatter of \xm \ was calculated instead of \xh, and all 12 clusters were combined to provide one single error estimate.  

During the full survey, MWM will observe a few tens of thousands of wide binaries, pairs of stars thought to have common origins \citep{2018MNRAS.473.5393A, 2019ApJ...871...42A, 2020MNRAS.492.1164H}. These are resolved on the sky, so that independent data can be obtained for both components. Given the common origin of their components, wide binaries can also be used for calibration purposes or to test the precision of our abundances. The pairs in the MWM wide binary program are selected from the catalogs of \citet{2021MNRAS.506.2269E} and \citet{2020ApJS..247...66H}.

The initial list of 950 binaries with observations in DR19 was down-selected by restricting the radial velocity difference of the primary and secondary to be within $\pm$ 5 \kms, the temperature of both components to be between 4000 and 6000~K, and SNR to be greater than 50. The final sample contained 348 wide binaries as summarized in Table~\ref{tab:validation-set}.  The scatter of the \xm \ abundance differences between the primary and secondary was calculated, then divided by the square root of two to estimate the precision of a single measurement. Unlike in the solar neighborhood and open cluster samples, the wide binaries contain both giant and main-sequence stars, as well as moderately metal-poor stars. 

In DR19, the precision of abundances is determined from multiple observations of the same stars and is explained in detail by Casey et al. (2025, in preparation). This uncertainty is calculated for \xh \ and not for \xm, and their values are stored in the e$\_$x$\_$h nametag for each zero-point calibrated abundance, and e$\_$raw$\_$x$\_$h nametag for the raw values. The two sets of uncertainties are different, and here, we discuss the uncertainty of the former for giant stars with \logg \ $<$ 3.8. We compare the ASPCAP uncertainty of \xh \ with independent uncertainty estimates of \xm, because the contribution of metallicity must be taken into account. However, Astra did not calculate the uncertainty of \xm.

The comparison of the three different types of uncertainty estimates and the median internal uncertainty as reported in Astra are shown in Figure~\ref{abuerrors} and listed in Table~\ref{abuerrorstab}. We discuss the discrepancy between these uncertainty estimates for each element separately in the subsequent element subsections. It is worth noting that the three independent uncertainty values may be overestimated, as there may be potential inhomogenities in the metallicities and abundances that have astrophysical origin, and the temperature-based corrections were not removed either from the overall scatter values. While we compare the individual values below for each element separately, it can be concluded that the median \xh \ uncertainties are all underestimated in DR19, and the estimates from the three independent sources serve as an upper limit, but still provide an easy assessment of the overall precision.

\begin{deluxetable}{lrrrr}
\tabletypesize{\scriptsize}
\tablecaption{Estimated Uncertainties of Abundances.}
\tablewidth{0pt}
\tablehead{
\colhead{Element} & \colhead{DR19} & 
\colhead{Solar} & \colhead{Open} & \colhead{Wide} \\
\colhead{} & \colhead{Median} & 
\colhead{Neighborhood} & \colhead{Clusters} & \colhead{Binaries} \\ 
\colhead{} & \colhead{\xh} & 
\colhead{\xm} & \colhead{\xm} & \colhead{\xm}
}
\startdata
$\alpha$         	& 0.003  &  0.012  & 0.027  & 0.019  \\
C$_{\rm global}$    & 0.005  &  0.060  & 0.083  & 0.047  \\
C$_{\rm windows}$   & 0.004  &  0.071  & 0.087  & 0.069  \\
C$_{\rm iso}$       & 0.004  &  0.217  & 0.188  & 0.279  \\
N$_{\rm global}$    & 0.008  &  0.113  & 0.084  & 0.098  \\
N$_{\rm windows}$   & 0.004  &  0.128  & 0.087  & 0.138  \\
O                   & 0.003  &  0.056  & 0.059  & 0.079  \\
Na                  & 0.001  &  0.167  & 0.099  & 0.434  \\
Mg                  & 0.003  &  0.053  & 0.041  & 0.040  \\
Al                  & 0.001  &  0.049  & 0.048  & 0.065  \\
Si                  & 0.003  &  0.023  & 0.027  & 0.032  \\
P                   & 0.001  &  0.269  & 0.248  & 0.330 \\
S                   & 0.003  &  0.067  & 0.089  & 0.151  \\
K                   & 0.001  &  0.093  & 0.081  & 0.106  \\
Ca                  & 0.003  &  0.034  & 0.036  & 0.059  \\
Ti                  & 0.003  &  0.082  & 0.073  & 0.149  \\
V                   & 0.002  &  0.265  & 0.272  & 0.479  \\
Cr                  & 0.001  &  0.078  & 0.077  & 0.178  \\
Mn                  & 0.002  &  0.035  & 0.054  & 0.077  \\
Co                  & 0.001  &  0.144  & 0.203  & 0.393   \\
Ni                  & 0.001  &  0.027  & 0.034  & 0.038  \\
Cu                  & 0.001  &  0.246  & 0.207  & 0.357  \\
Ce                  & 0.001  &  0.231  & 0.225  & 0.231  \\
Nd                  & 0.001  &  0.149  & 0.223  & 0.125  \\
\enddata
\label{abuerrorstab}
\end{deluxetable}

\subsection{The Overall Quality of the ASPCAP Abundances in DR19}
\label{sect:quality_abund}

Before describing each element in detail, we organize them into four categories according to their overall quality listed in Table~\ref{tab:abuerrorstab}. This is done to provide a summary of information for the users for their science applications. 

To reach the final conclusion of Table~\ref{tab:abuerrorstab}, we quantified the categories excellent, good, fair, and poor based on three criteria, as summarized in Table~\ref{tab:abuerrorstab}. These criteria are based on the discussions in previous sections. More specifically, for accuracy (first column), we follow Section~\ref{accuabusec} and assign the accuracy quality criterion using the value of zero-point offset ($\Delta$) of each abundance defined using the SNSM sample. These values are listed in Table~\ref{zeropoint}. The systematics (second column) are examined through the dependencies of the abundances as a function of \teff \ in the open cluster sample (Section~\ref{abuteffsec}). Using the $a$ coefficients of Table~\ref{corrcoeff}, we can calculate the effective difference in predicted abundances, $\delta$, between the edges of our temperature range, namely  \teff = 6000 and \teff= 4000 K. We can thus define the ranges for the quality of the abundance due to systematics in this temperature range, which are listed in the sixth column of Table~\ref{corrcoeff}. The last quantified criterion is precision, shown in the third column of Table~\ref{tab:abuerrorstab}, and follows the analysis discussed in Section~\ref{precabu}, summarized in Table~\ref{abuerrorstab}. In addition, we consider a fourth criterion related to the range in the parameter space that an element has well-measured abundances. This is discussed in detail for each element below, and also helps in assessing the overall quality of each element. The last column in Table~\ref{abuerrorstab} indicates the elements that fall in each category based on these quality criteria and the detailed discussion of each individual element in Section~\ref{elements}. 

There are 10 elements (M, $\alpha$, C, N, O, Mg, Si, Ca, Fe, and Ni) that we consider to have excellent quality in DR19. These are the elements that have small zero-point offsets, no significant dependencies with temperature, very high precision, and can be used in a large parameter space. In the second category, the abundances of the five elements Al, S, K, Cr, and Mn are considered good, because they have somewhat larger zero-point offsets and poorer precision, or show trends with temperature in the open cluster and/or solar neighborhood sample. The abundances of Na, Ti, Co, Ce, and Nd are only of fair quality in DR19, because we were only able to measure them reliably in a narrow parameter region, usually below 6000~K and for solar-metallicity giant stars. The absorption lines of these elements are usually quite weak, which prevents better measurements. In the final category lie P, V, Cu, which are of poor quality, and unfortunately, our measurements of these elements are not reliable in any of the observed parameter space.

\begin{figure*}                          
\centering
\includegraphics[width=7.1in,angle=0]{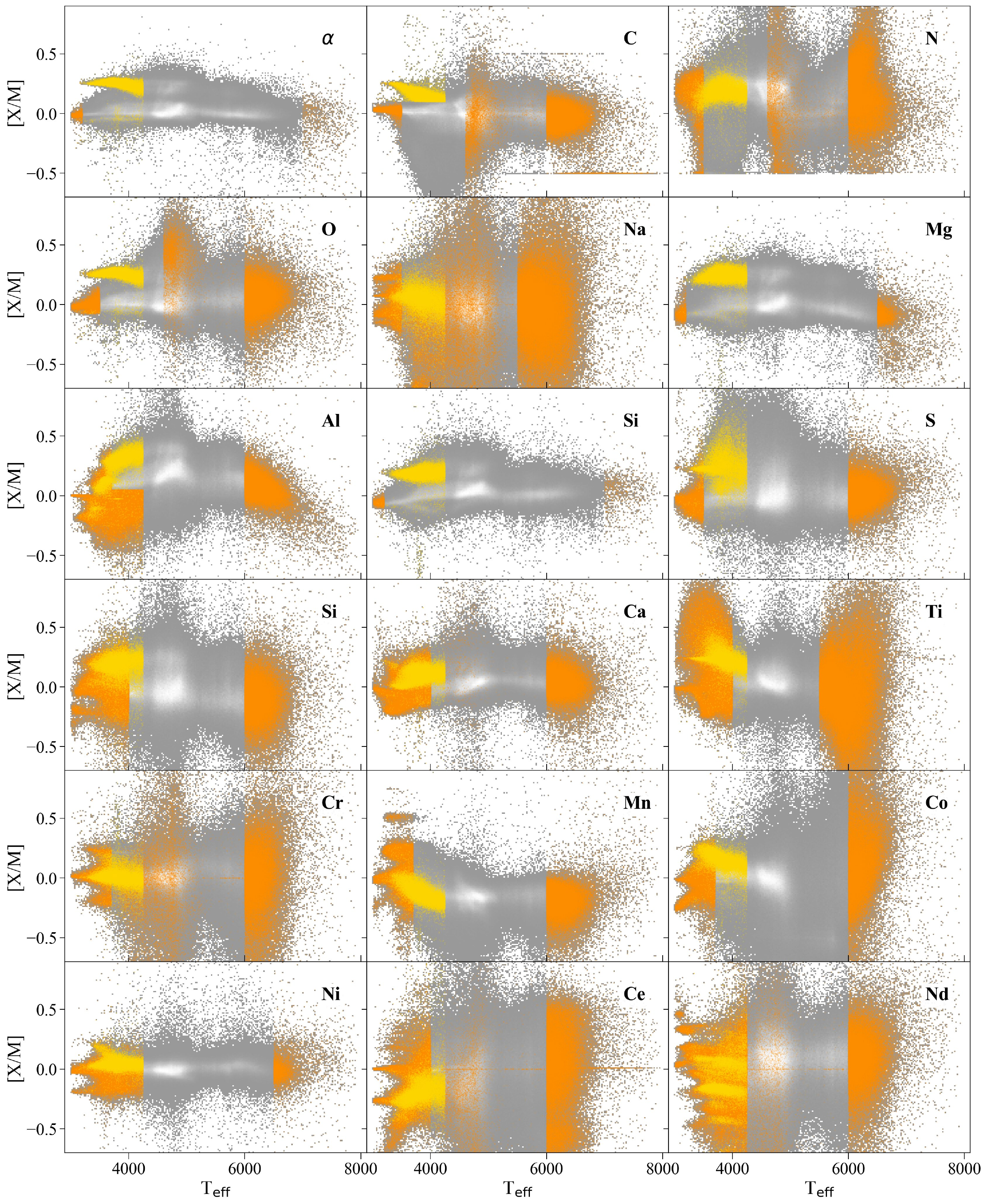}
\caption{Raw abundances of giants (\logg \ $<$ 3.8) as a function of \teff \ with SNR $>$ 50 and no BAD flags. The orange coloring indicates the parameter region from Table~\ref{abuerrorstab2} where the abundances might not be reliable. The yellow region identifies the cool and carbon-rich stars with \teff \ $<$ 4250~K and \cm \ $>$ 0.1.}
\label{abuflags}
\end{figure*}

The detailed assessment of each of these three criteria for all elements is listed in Table~\ref{abuerrorstab2} and shown in Figure\ref{abuflags}. This table also contains the parameter range, separated for giants and dwarfs in which most of the given abundance of choice is reliable. The listed ranges act as a guide for users, but are not necessarily designed to be used as a strict parameter cut to be implemented for all science applications. The reason behind this is that atomic and molecular lines gradually weaken with increasing temperature \citep{2005oasp.book.....G}. This usually results in increasing uncertainty above 6000K, but not necessarily incorrect measurements. For the same reason, molecular lines gradually overtake the spectrum at temperatures below 4000~K, making it difficult to accurately fit individual atomic absorption lines \citep{2012A&A...547A.108L, 2019ARA&A..57..571J}. Users are advised to carefully check abundances outside these parameter ranges and refine their parameter cuts according to the need of their science application. Thus, our main motivation with Table~\ref{abuerrorstab2} is to identify problematic regions in \teff, \mh, \logg \ space where abundances have high uncertainty, and/or show signs on clumping around one single value that does not have an astrophysical origin.

One such region that affects all abundances in DR19 is a group of stars that can be defined as having \teff \ $<$ 4250~K and \cm \ $>$ 0.1~dex. These stars can be seen in Figure\ref{fig:carb} and subsequently denoted by yellow dots for all elements. This high carbon group also has an elevated $\alpha$ abundance, high individual $\alpha$ elements and metallicity higher than $-$1; however, based on their distribution in the Milky Way, they are not part of the thick disk. Thus, we believe the existence of this group signals the inability of ASPCAP to accurately fit the multidimensional APSCAP grid to the spectra of these stars, although the exact reasons are currently unknown.

As established in earlier sections, the main atmospheric parameters (\teff, \logg, \mh) of main-sequence stars below \teff \ $<$ 4500~K have significant offsets from independent measurements. This systematic error may also propagate into the measurement of individual abundances; thus, users should use the abundances of these dwarf stars with caution, even if we consider them to have excellent quality in general. As discussed in Section~\ref{sec:metallicity} some of these cool dwarf stars exhibit unreasonably low metallicities, and these wrong measurements also propagate into individual \xm \ values.

The medians of the uncertainties reported by Astra are listed in Table~\ref{abuerrorstab}. These values are generally very low, varying between 0.001 and 0.008~dex depending on the element. These values are derived from repeated observations of the same stars, and suggest excellent stability of the APOGEE instrument, reduction pipeline, and ASPCAP. These uncertainties are compared to the overall uncertainty estimated from the scatter of abundances in open clusters, SNSM sample, and wide binaries in Table~\ref{abuerrorstab}. Even though those estimates may only serve as an upper limit, we can still conclude that the internally reported uncertainties are generally several times lower than what is estimated from independent sources.


\section{Discussion of Individual Abundances}
\label{elements}

\subsection{$\alpha$ Elements}

\begin{figure*}                          
\centering
\includegraphics[width=7.1in,angle=0]{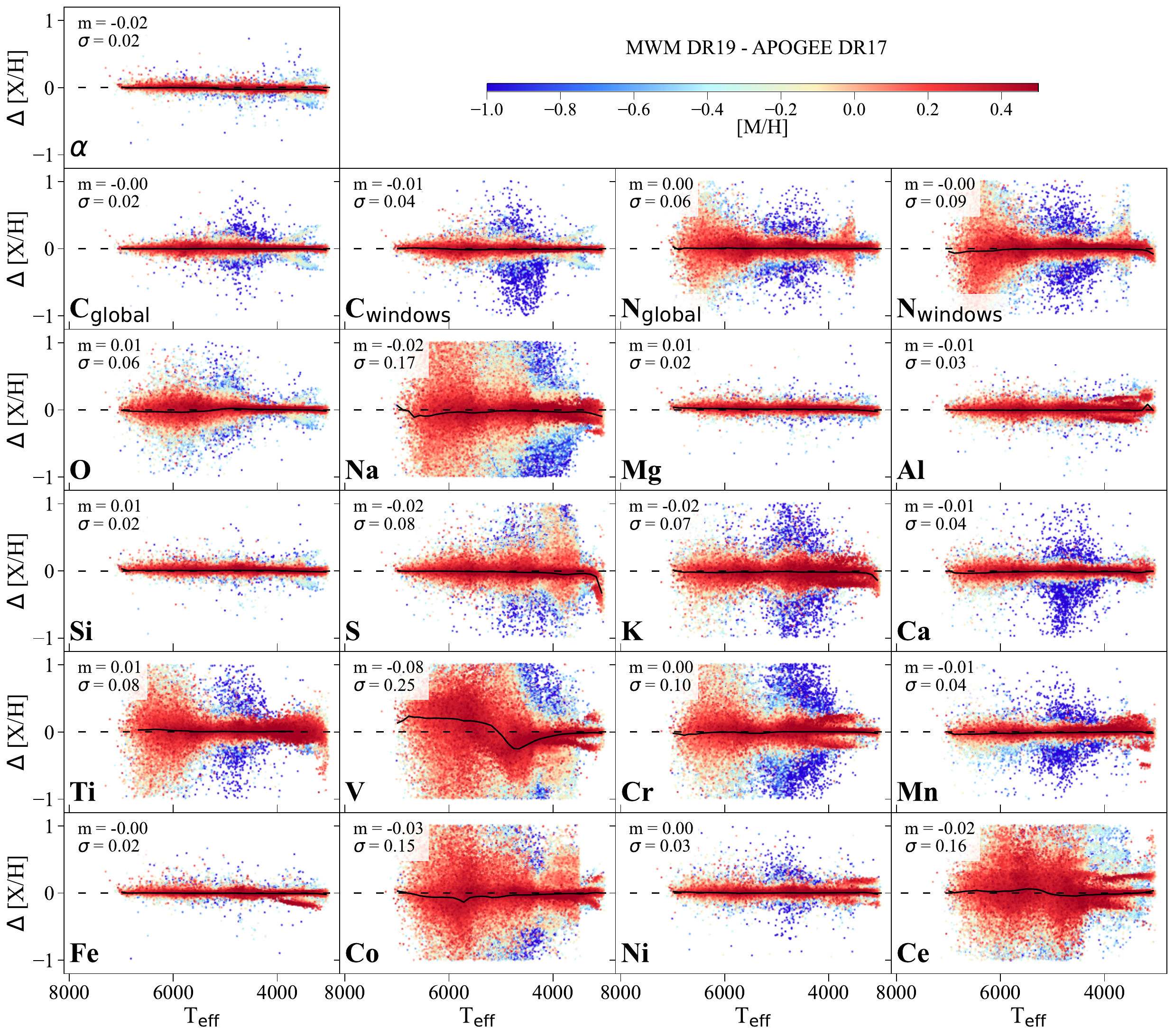}
\caption{Comparison between MWM DR19 and APOGEE DR17 abundances as a function of \teff \ color coded by MWM \mh. \snr \ was restricted to be larger than 50 in both surveys. Abundances of C$_{\rm global}$ and N$_{\rm global}$ are from the global fit. The solid black line shows the running median value.}
\label{mwmapogee}
\end{figure*}

The $\alpha$ \ elements (raw\_a\_m\_atm nametag) considered in ASPCAP are the following: O, Mg, Si, S, Ca, and Ti. When deriving  \am, ASPCAP uses a wavelength window sensitive to all of the  absorption lines of these elements. Thus, \am \ reflects the abundance of O, Mg, and Si because these elements have the strongest lines, while S, Ca, and Ti have lower weights than the other three elements. The \am \ vs. \teff \ show a near-constant value in the SNSM sample (Figure~\ref{solarneigh}) for both giants and dwarfs; any discrepancies shown are within our uncertainties. This constant offset suggests that \am \ values are very accurate to at least 3500~K, and also above 7000~K (Table~\ref{abuerrorstab2}), though the scatter noticeably increases above this temperature. The agreement with APOGEE DR17 $\alpha$ abundances is also excellent (Figure~\ref{mwmapogee}).

\begin{figure}                          
\centering
\includegraphics[width=3.42in,angle=0]{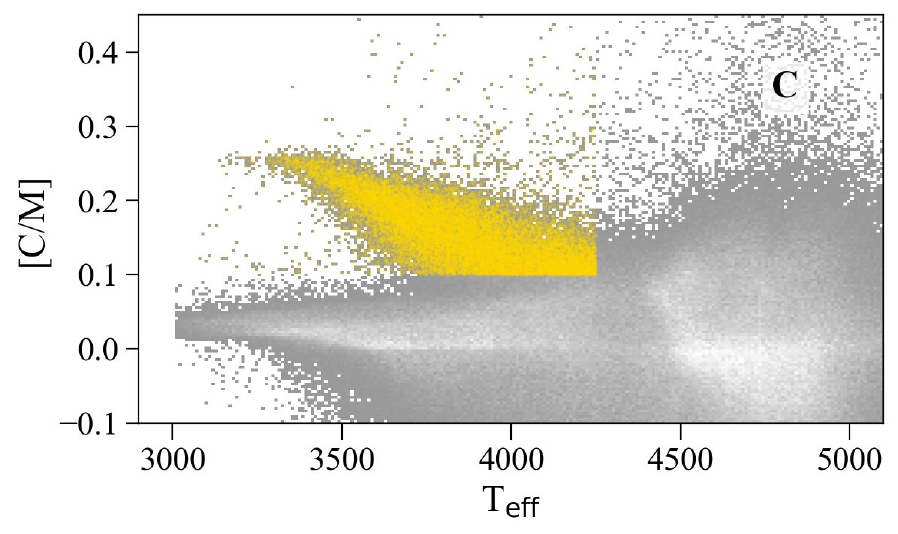}
\caption{\cm \ as a function of \teff \ below 5000~K for the entire sample with SNR $>$ 50 and no BAD flags. The group of stars with \teff \ $<$ 4250~K and \cm \ $>$ 0.1 are shown with yellow dots.}
\label{fig:carb}
\end{figure}

The elevated $\alpha$ values of the thick disk and the near solar \am \ of the thin disk separate easily in Figure~\ref{abunmap}, with no obvious noding visible. An unexpected grouping of high$-\alpha$ stars can be seen in Figure\ref{abuflags} below 4250~K, which is the group of cool stars that is also rich in carbon and are probably the result of incorrect fitting of stellar spectra. Other than this issue, \am \ is the most-precise abundance MWM measured in DR19. Based on our three independent data sets described in Section~\ref{precabu} the estimated uncertainty varies between 0.012 and 0.027~dex (Figure~\ref{abuerrors} and Table~\ref{abuerrorstab}). Overall, both the accuracy, precision, and systematics of \am \ are excellent in DR19.

\begin{deluxetable*}{lcccccc}
\tabletypesize{\scriptsize}
\tablecaption{Overall Assessment of the Quality of abundances published in DR19.
}
\tablewidth{0pt}
\tablehead{
\colhead{} & \colhead{Accuracy} & 
\colhead{Systematics}  &
\colhead{Precision} & 
\colhead{Final Assessment of Elements} \\
\colhead{} & \colhead{$\Delta$} & 
\colhead{$\delta$}  &
\colhead{$\sigma$} & 
\colhead{}
}
\startdata
Excellent  & $<0.05$ & $< 0.1$ & $<0.1$ & M, $\alpha$, C, N, O, Mg, Si, Ca, Fe, Ni \\
Good & $[0.05, 0.1]$ & $[0.1, 0.15]$ & $[0.1, 0.15]$ & Al, S, K, Cr, Mn \\
Fair & $[0.1, 0.15]$ & $[0.15, 0.25]$ & $[0.15, 0.3]$& Na, Ti, Co, Ce, Nd \\
Poor & $>0.15$ & $>0.25$ & $>0.3$& P, V, Cu \\
\enddata
\tablecomments{Details can be found in Sect.\ref{sect:quality_abund}}
\label{tab:abuerrorstab}
\end{deluxetable*}

\subsection{Carbon}

There are two different carbon abundances published in DR19: \cm \ from the global fit (raw\_c\_m\_atm nametag) and \ch \ from the CO molecular line windows (raw\_c\_h). The abundance of carbon can also be derived from the very weak 16 895 \AA \ absorption line of the neutral C atom, but its value is not published in DR19, because its derivation was not successful. Since the strength of CO lines also depends on the O abundance, minor degeneracies between the two abundance are possible, although strong evidence of their existence is not present in the DR19 data. The carbon abundance as determined from molecular lines is more precise than the carbon abundance determined from the C I lines, because the atomic lines are generally weak in the $H$ band. 

As mentioned in Sec.~\ref{sect:quality_abund} cool stars with \teff \ $<$ 4250~K can have unreasonably high \cm \ values (from the global fit) above 0.1~dex regardless of surface gravity. This region is shown in Figure~\ref{fig:carb} in detail. Users are advised to handle all abundances of these stars with care, because carbon abundances affect all spectral regions through CO and CN lines, and \cm \ is determined from the global fit of the spectra and then fixed in the subsequent phase of determining  individual abundances. These stars are also denoted by yellow dots in Figure\ref{abuflags}.

Because material that has been processed through the CNO cycle in the main sequence is brought to the surface when the star ascends the RGB, surface carbon abundances will be depleted, while nitrogen will be enhanced \citep[][and references therein]{2019A&A...621A..24L}. For this reason, C measurements do not have zero-point, or \teff-dependent calibrations, and the scatter measured in the solar neighborhood, open clusters, or wide binaries only give an upper estimate of our precision. These scatters vary between 0.047 and 0.087~dex for \cm \ and \ch \, which means that we measure \cm \ and \ch \ with excellent precision, as those scatter values also contain the effect of mixing. Carbon abundances from the global fit and the CO molecular lines have very similar uncertainties (Table~\ref{abuerrorstab}). However, there appears to be a small offset of about 0.01$-$0.06~dex between the two carbon scales for the thin disk stars (see the carbon panels in Figure~\ref{abunmap}). The agreement between MWM and GALAH DR4 carbon abundances is acceptable (Figure~\ref{mwmgalah}), with an overall scatter in the differences smaller than 0.18~dex. The metallicity dependent discrepancies visible below 5000~K may be the result of the differences in metallicities between these two surveys detailed in Section~\ref{sec:metallicity}.

The strength of CO lines varies with temperature and metallicity, and when \mh \ is below $-$1~dex, the CO lines become insensitive to amount of carbon in the atmosphere above 4500$-$4600~K \citep{2015AJ....149..153M, 2019A&A...622A.191M}. This is why there is a larger discrepancy between the carbon abundances of MWM DR19 and APOGEE DR17 of low-metallicity stars (see carbon panels of Figure~\ref{mwmapogee}), than at solar metallicities. At solar metallicity, CO lines are stronger and can be used to measure \ch \ up to about 5500$-$6000~K (Table~\ref{abuerrorstab2}). Consequently, carbon abundances derived outside these regions should be treated with caution. Considering these limits, the quality of \cm \ from the global fit and CO molecular lines is excellent. 

${^{12}}{}{\rm C} / {^{13}}{}{\rm C}$ isotopic ratios (raw\_c\_12\_13) are also derived part of DR19 by measuring the abundance of ${^{13}}{}{\rm CO}$ lines. These abundances are unfortunately of poor quality (scatter changes between 0.188 and 0.279~dex in our reference samples) in DR19 and users advised to carefully check the values before any science application. However, with careful manual analysis, it is possible to measure reliable ${^{12}}{}{\rm C} / {^{13}}{}{\rm C}$ isotopic ratios as presented by \citet{2018MNRAS.474.4810S, 2022ApJS..262...34H}.

\subsection{Nitrogen}

There are two nitrogen abundances published in DR19: \nm \ from the global fit (raw\_n\_m\_atm), and \nh \ from fitting the CN molecular lines (raw\_n\_h), which could result in minor degeneracies with the carbon abundance. As nitrogen is also affected by deep mixing, no calibrations were applied to these two abundances. The scatter of \nm \ from the global fit is about 0.084$-$0.113~dex, while from the windows, it varies between 0.087 and 0.138~dex (Table~\ref{abuerrorstab} and Figure~\ref{abuerrors}), but this is mostly driven by mixing effects along the RGB. CN lines cover most of the $H$ band, not just the most sensitive wavelength windows selected for \nh \, which might explain why the global fit provides slightly better precision. Nevertheless, it appears that we measure \nm \ with close to excellent precision, probably slightly better than 0.1~dex in most cases.

The nitrogen abundances from the global fit of thin disk stars are slightly elevated by about 0.01$-$0.04~dex compared to the nitrogen abundance from the windows; see the nitrogen panels in Figure~\ref{abunmap}. Similarly to CO, the strength of CN lines becomes insensitive to nitrogen for stars with \mh \ is below $-$1~dex, and above 4500$-$4600~K \citep{2015AJ....149..153M, 2019A&A...622A.191M}. At solar metallicity, CN lines cannot be used to measure \nh \ above about 6000~K. The comparison with APOGEE DR17 reveals good agreement, except above 6000~K for all metallicities, and the large scatter seen in metal-poor stars (Figure~\ref{mwmapogee}). Nitrogen abundances derived from these regions have excellent quality, but outside should be treated with caution similarly to carbon and oxygen. There is a significant correlation visible between the difference of DR19 and GALAH DR4 nitrogen abundances and \teff \  (Figure~\ref{mwmgalah}).

\subsection{Oxygen}

The oxygen abundance (raw\_o\_h) is derived from OH molecular lines in the $H$ band, and has minor zero-point offsets of $-$0.004 and 0.024~dex for giants and dwarfs, respectively. Surface oxygen abundances are not affected by mixing, because O remains constant during the CNO cycle; thus, we can use the SNSM  stars to assess how the accuracy and precision varies with temperature. Based on the O panels in Figure~\ref{solarneigh}, we conclude that OH lines can be reliably measured to about 6000~K, but above this temperature, the scatter of O significantly increases because OH lines start to disappear from the spectra above that temperature. Although there are very few stars below 4000~K in this sample, those \om \ values are close to solar, suggesting that we can reliably measure \om \ even down to 3500~K (Table~\ref{abuerrorstab2}). Below this temperature, strong blends generally make it difficult to fit any absorption lines. There are not any unexpected features or grouping of stars in the O map of the Milky Way (see the O panel in Figure~\ref{abunmap}), as both the thin and thick disk nicely separate from each other. For dwarfs, \om \ values correlate with \teff, but it is still fairly close to the expected solar values with increasing scatter as temperature rises, similarly to giants. 

At low metallicities, the reliability of measuring \oh \ decreases. The strength of the OH lines become insensitive to oxygen for stars with \mh \ below $-$1~dex, and above 4500$-$4600~K \citep{2019A&A...622A.191M}. Oxygen abundances derived in these regions should be treated with caution. But even below this temperature, there are special cases where the oxygen abundance is not reliable. Because O is tied with the $\alpha$ dimension, ASPCAP is not capable of reliably fitting the OH lines when the [O/$\alpha$] is large. This happens mostly in globular clusters, where second-generation stars are oxygen poor, but rich in other $\alpha$ elements. \citet{2015AJ....149..153M, 2019A&A...622A.191M, 2020MNRAS.492.1641M} provide more discussion of this issue.

The differences from APOGEE DR17 increase at low metallicities and above about 5000~K, but the general agreement is very good (Figure~\ref{mwmapogee}). The scatter of \om \ in our three independent reference samples varies between 0.056 and 0.079~dex (Table~\ref{abuerrorstab} and Figure~\ref{abuerrors}), thus oxygen is one of the most-precise elements published in DR19. Considering the excellent precision of \oh \, and how sensitive O is to \teff, the relatively large scatter of 0.25~dex seen in the MWM-GALAH common sample (see the O panel in Figure~\ref{mwmgalah}) probably originates from the \teff \ and \mh \ differences between the two surveys. Overall, the quality of \om \ in DR19 within the above listed regions is considered excellent.

\subsection{Sodium}

Sodium (raw\_na\_h) is one of the least-precise elements published in DR19, because its two fairly weak lines in the $H$ band can only be measured in the spectra of metal-rich stars, explaining the large scatter seen in the DR19 vs. APOGEE DR17 comparison (Figure~\ref{mwmapogee}). The measured scatter is 0.167~dex in the SNSM sample, and 0.099~dex in open clusters. These two samples contain only giant stars, but the 0.434~dex scatter in wide binaries indicates the Na abundances in dwarfs are mostly unreliable, which is reinforced by the large scatter seen in the dwarf panel of Na in Figure~\ref{solarneigh}. 

Based on the scatter seen in the giants of the SNSM sample, it becomes increasingly difficult to fit the Na lines above 5500~K and below 3500~K the Na abundances become supersolar. For dwarfs, this lower limit is larger around 4200$-$4300~K, below which multiple blobs are visible in the Na panel of Figure~\ref{solarneigh2}. The strength of Na lines becomes close to indistinguishable from the continuum, even in the spectra of giants roughly below \mh \ $< -0.7 - -1.0$~dex. Thus, measuring the Na abundance is possible only near solar metallicity. The MWM-GALAH sample also shows a large scatter (Figure~\ref{mwmgalah}), which is mainly due to the low precision of our measurements. Taking into account these limitations (Table~\ref{abuerrorstab2}) the overall quality of the Na abundance in DR19 is only fair.

\subsection{Magnesium}

Magnesium (raw\_mg\_h) is one of the most-precise elements published in DR19. The zero-point offset is negligible in giants, but $-$0.031 for dwarfs, which is still quite small. The agreement between DR19 and APOGEE DR17 is excellent (Figure~\ref{mwmapogee}). Similarly to other elements, the scatter of Mg for giants increases above 6500~K and the abundance values become slightly subsolar (Figure~\ref{solarneigh}). \mgh \ appears to depend on \teff \ in the dwarf stars of the SNSM sample, suggesting that measuring Mg in the spectra of main-sequence stars is less accurate than that of giants. Nevertheless, \mgh \ values are close to 0 even below 4000~K suggesting that Mg abundances are reliable in the coolest stars down to about 3200~K for giants and 3500~K for dwarfs (Table~\ref{abuerrorstab2}).

The chemical map of Mg in Figure~\ref{abunmap} shows the standard separation of the thin and thick disks; however, there is a group of stars with slightly sub-solar metallicity that have \mgm \ content between $-$0.2 and $-$0.5~dex (some of these are also O-poor). These are mostly dwarfs that may have incorrect Mg abundances. The estimated precision varies between 0.04 and 0.053~dex in our test samples. The agreement between MWM and GALAH Mg abundances (Figure~\ref{mwmgalah}) is also very good with no significant offset and relatively low scatter of 0.11~dex, which is dominated by the larger scatter seen below 4500~K due to the metallicity differences between the two surveys (see Section~\ref{sec:metallicity} for the explanation of this issue). Thus, the quality of \mgh \ is all around excellent in DR19.

\begin{deluxetable*}{lrrrcc}
\tabletypesize{\scriptsize}
\tablecaption{Quality of Individual Abundances
}
\tablewidth{0pt}
\tablehead{
\colhead{Element} & \colhead{Accuracy} & 
\colhead{Systematics}  &
\colhead{Precision} & 
\colhead{Optimal Region} & 
\colhead{Optimal Region} \\
\colhead{} & \colhead{} & 
\colhead{}  &
\colhead{} & 
\colhead{\logg \ $<$ 3.8} & 
\colhead{\logg \ $\geq$ 3.8}}
\startdata
$\alpha$ & excellent  &  excellent  &  excellent  & 7000$-$3200K & 7000$-$3500K \\
C$_{\rm global}$  & \nodata  & \nodata & excellent & 4600$-$3500K, \mh \ $<-$1  & 6000$-$3500K, \mh \ $ >-$0.5 \\
  &   &             &             & 6000$-$3500K, \mh \ $>-$0.5  &  \\
C$_{\rm windows}$ & \nodata  & \nodata & excellent & 4600$-$3500K, \mh \ $<-$1 & 6000$-$3500K, \mh \ $ >-$0.5 \\
&   &             &               &  6000$-$3500K, \mh \ $>-$0.5  &  \\
C$_{\rm iso}$     & \nodata & \nodata & poor & nowhere & nowhere \\
N$_{\rm global}$  & \nodata       & \nodata & good/excellent & 4600$-$3500K, \mh \ $<-$1 & nowhere \\
  &        &             &         & 6000$-$3500K, \mh \ $>-$0.5  & nowhere \\
N$_{\rm windows}$ & \nodata      & \nodata & good/excellent & 4600$-$3500K, \mh \ $<-$1 & nowhere \\
                  &         & &    & 6000$-$3500K, \mh \ $>-$0.5  & nowhere \\
O   & excellent  &  excellent  &  excellent  & 4600$-$3500K, \mh \ $<-$1  & 4600$-$3500K, \mh \ $<-$1 \\
 &    &     &     &                 6000$-$3500K, \mh \ $>-$0.5    &  6000$-$3500K, \mh \ $>-$0.5 ? \\

Na  & good       &  fair  &  poor/fair       & 5500$-$3500K, \mh \ $>-$0.7 & 5500$-$4250K, \mh \ $>-$0.7 \\
Mg  & excellent  &  excellent  &  excellent  & 6500$-$3200K & 7000$-$3500K \\
Al  & poor  &  excellent  &  excellent  & 6000$-$4250K & 7000$-$4250K \\
  &   &         &                            & 4250$-$3700K, \alm \ $>$ 0.05 &  \\
Si  & excellent  &  excellent  &  excellent  & 7000$-$3200K & 7000$-$3500K? \\
P   & fair       &  poor       &  poor/fair       & nowhere   & nowhere \\
S   & excellent       &  excellent  &  good/excellent       & 6000$-$3500K  & 6000$-$4500K \\
K   & good       &  good       &  good/excellent       & 6000$-$4000K  & 6500$-$3500K \\
Ca  & excellent  &  good &  excellent        & 6000$-$4000K, \mh \ $>-$2.0 & 7000$-$3500K, \mh \ $> -$2.0 \\
Ti  & excellent       &  fair       &  good/excellent       & 5500$-$4000K & 5500$-$4250K \\
V   & fair       &  poor       &  poor/fair       & nowhere  & nowhere \\
Cr  & excellent  &  good       &  good/excellent  & 6000$-$3700K, \mh \ $>-$1.0  & 6000$-$4250K, \mh \ $>-$1.0 \\
Mn  & fair       &  good  &  excellent       & 6000$-$3700K & 7000$-$4000K \\
Co  & good       &  poor       &  poor/fair  & 6000$-$3700K  & nowhere  \\
Ni  & excellent  &  excellent  &  excellent  & 6500$-$4250K & 7000$-$3500K \\
  &   &         &                            & 4250$-$3700K, \nim \ $>-$0.1 &  \\
Cu  & poor       &  poor    &  poor/fair  & nowhere    & nowhere \\
Ce  & fair       &  poor       &  fair       & 6000$-$4000K, \mh \ $>-$1.5  & nowhere \\
Nd  & fair       &  good       &  fair/good       & 6000$-$4250K, \mh \ $>-$1.5 & 5500$-$4250K?, \mh \ $>-$1.5 \\
\enddata
\label{abuerrorstab2}
\end{deluxetable*}

\subsection{Aluminum}

Aluminum abundance (raw\_al\_h) is derived from three very strong unblended lines in the $H$ band; however, relatively large zero-point offsets of 0.175 and 0.075~dex for giants and dwarfs (Table~\ref{zeropoint}), respectively, were necessary to force \am \ to be solar based on the SNSM sample. One possible explanation could be NLTE effects, which were not taken into account in the spectral grid in DR19. The lack of NLTE correction may cause the temperate dependence of \alh \ in this sample, which is more complicated than a simple constant offset. This is illustrated in the Al panel of Figure~\ref{solarneigh}, where the corrected \alh \ of both giants and dwarfs strongly correlates with \teff, and exhibits a bimodality in the dwarf sample below about 4200$-$4300~K. Nevertheless, Al lines are still sensitive to the Al content in the atmosphere even at very low metallicities down to $-$2.5~dex and high temperatures up to roughly 6000~K for giants and 7000~K for dwarfs (Table~\ref{abuerrorstab2}), but obviously precision suffers outside these parameter regions.

\begin{figure*}                          
\centering
\includegraphics[width=7.1in,angle=0]{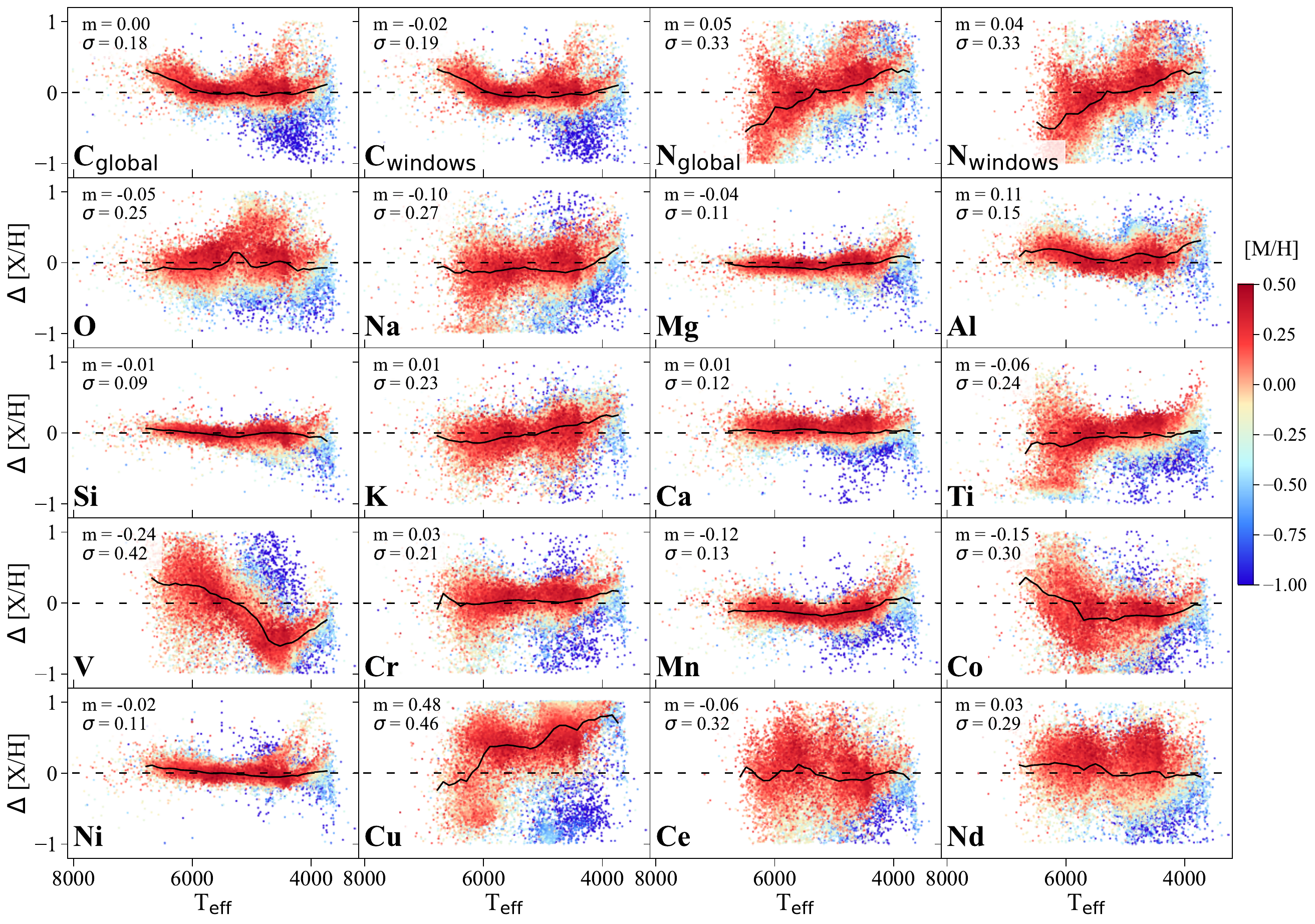}
\caption{Comparison between MWM DR19 and GALAH DR4
abundances as a function of \teff \ color coded by MWM \mh. \snr \ was restricted to be larger than 50 in both surveys. Abundances of C$_{\rm global}$ and N$_{\rm global}$ are from the global fit. The solid black line shows the running median value.}
\label{mwmgalah}
\end{figure*}

Our estimated precision of Al is comparable to that of Mg, varying between 0.048 and 0.065~dex (Table~\ref{abuerrorstab} and Figure~\ref{abuerrors}). While the precision is excellent, not all Al abundances seem to be correct in the full DR19 sample. There is a group of stars with \alm \ between lower than $-$0.1 and $-$0.4~dex between \mh \ = $-$0.5 and 0.5~dex, visible in the Al panel of Figure~\ref{abunmap}, and multiple stripes are visible below \alm \ $<$ 0.05 and \teff \ $<$ 4250~K (Figure~\ref{abuflags}), that may not have an astrophysical origin. These stripes also appear in comparison with APOGEE DR17 Al abundances (Figure~\ref{mwmapogee}). Considering all these limitations, Al abundances are still reliable in a large parameter space listed in Table~\ref{abuerrorstab2}. The agreement of \alh \ between MWM and GALAH (Figure~\ref{mwmgalah}) is generally good, though there appears to be a small \teff \ dependence and a median offset of 0.11~dex. Considering the poor accuracy of \alm \ and the existence of multiple blobs and stripes at low temperatures, but the excellent precision and systematics, the overall quality of the Al abundance in DR19 is good (Table~\ref{tab:abuerrorstab}).

\subsection{Silicon}

Similarly to $\alpha$, O, and Mg, the abundance of Si (raw\_si\_h) published in DR19 is among the most precise, better than 0.032~dex in all of our estimates (Table~\ref{abuerrorstab} and Figure~\ref{abuerrors}). The agreement between DR19 and APOGEE DR17 abundances of Si is excellent (Figure~\ref{mwmapogee}). The accuracy is similarly excellent for both giants and dwarfs with low zero-point offsets of 0.023 and $-$0.030~dex, respectively (Table~\ref{zeropoint}). Looking at the temperature dependence of \mgh \ in the SNSM sample, we conclude that Si abundances are reliable up to about 7000~K with some temperature dependence visible for dwarfs below 3500~K (Table~\ref{abuerrorstab2}). Similarly to O and Mg, the thin and thick disk separate well based on Si (Figure~\ref{abunmap}). The difference between MWM and GALAH \sih \ values has a small scatter of 0.09~dex with no signs of temperature dependence (Figure~\ref{mwmgalah}), this \sih \ has excellent quality in DR19.

\subsection{Phosphorous}

Because phosphorous (raw\_p\_h) is measured from two very weak lines in the $H$ band, both its accuracy and precision suffer significantly to the point where we believe most of the \ph \ abundances in DR19 are unreliable. The poor precision can be seen in the SNSM stars in the P panel of Figure~\ref{solarneigh}, where both the giants and dwarfs show very large scatter. In fact, even under the most optimistic case, the precision of P is not better than 0.248~dex. There are also significant multimodalities present in the \ph \ $-$ \mh \ diagram (Figure~\ref{abunmap}), which most likely do not have an astrophysical origin. However, in special cases, \ph \ can be derived from MWM spectra for P-rich stars as demonstrated by \citet{2020NatCo..11.3759M}. Because \ph \ has a poor quality, we recommend the user to handle \ph \ with caution and check the best-fit spectra of the P lines before any scientific analysis. Deriving \ph \ from the $H$ band spectra is possible, and one such careful semi-automated analysis was presented by \citet{2022ApJS..262...34H} who derived P abundances for 120,000 stars using APOGEE data.

\subsection{Sulfur}

The abundance of sulfur (raw\_s\_h) is determined from two moderately weak lines; thus, it is less precise than other $\alpha$ elements like O, Mg, and Si. Our estimated uncertainties vary between 0.067 and 0.151~dex (Table~\ref{abuerrorstab} and Figure~\ref{abuerrors}). This moderate scatter is also visible in the DR19 and APOGEE DR17 common sample (Figure~\ref{mwmapogee}). The thick and thin disk stars are not well separated in the sulfur panel of Figure~\ref{abunmap}, either. The zero-point offsets of 0.044 (giants) and 0.005~dex (dwarfs) are relatively small, and no significant correlations with temperature can be seen in the SNSM sample, even up to 6000~K in case of dwarfs. For giants, the scatter significantly increases above 6000~K, signaling the disappearance of S lines in the spectra of giants above this temperature. On the lower side, \sh \ is probably reliable down to 3500~K for giants, but has a large scatter below 4500~K for dwarfs (Figure~\ref{solarneigh2}). Still, the overall quality of \sh \ in DR19 is generally good.

\subsection{Potassium}

Potassium abundances (raw$\_$k$\_$h) are derived from two relatively strong lines in the $H$ band and have good estimated precision between 0.081 and 0.106~dex in our three reference samples (Table~\ref{abuerrorstab} and Figure~\ref{abuerrors}). The zero-point offsets are also good: $-$0.066 and $-$0.110~dex for giants and dwarfs respectively (Table~\ref{zeropoint}). A significant temperature correlation is visible in the SNSM sample for the dwarfs (Figure~\ref{solarneigh}), suggesting that the dwarf K abundances are less reliable than giants. The \kh \ abundance becomes unreliable above 6000~k for giants and 6500~K for dwarfs as both K lines become too weak above this temperature. The scatter visible in the NWM DR19-APOGEE DR17 common sample is moderate for metal-rich stars, but significant below \mh \ $ < -$1~dex (Figure~\ref{mwmapogee}), suggesting decreased precision at low metallicities.

There is a group of stars above \mh \ $>$ 0.0~dex that have elevated \km \ abundance of about 0.3$-$0.5~dex, which is significantly higher than what is expected in this metallicity range in the thin disk (see the K panel of Figure~\ref{abunmap}). These stars are usually cool giants below 4000~K, which suggests that deriving K abundances for giant stars is challenging due to heavy blending of the K lines with other absorption features. Comparison with the optical GALAH K abundances does not show any systematic offset as a function of \teff \  (Figure~\ref{mwmgalah}). Users are advised to carefully examine the fits of K lines of cool giants with the above parameters before any science application, but the quality of potassium abundances in DR19 is generally good.

\subsection{Calcium}

Even though calcium abundances (raw$\_$ca$\_$h) are determined from moderately strong Ca lines, their overall quality is excellent, the estimated precision varies between 0.034 and 0.059~dex based on solar neighborhood stars, open clusters, and wide binaries (Table~\ref{abuerrorstab} and Figure~\ref{abuerrors}). The agreement with GALAH \cah \ values is also fairly good with a scatter of 0.12~dex and no \teff-dependent differences. There seems to be a slight metallicity dependence of \cah \ values when compared with GALAH DR4 below 5000~K. Although the zero-point offsets are larger than that of O, Mg, or Si (0.032 for giants and 0.063~dex for dwarfs; see Table~\ref{zeropoint}), there are no obvious correlations with \teff \ in the SNSM, the giants, or the dwarf sample. The scatter in the solar neighborhood increases significantly above 6000~K for giants (Figure~\ref{solarneigh}), because the Ca lines quickly become too weak to reliably fit above this temperature. This may explain the larger-than-expected scatter visible in the \cam \ vs. \feh \ distribution in the thin disk near solar metallicity (see in the Ca panel of Figure~\ref{abunmap}). Ca lines also start to disappear from spectra of stars below \mh \ $< -$2.0, so the precision of \cah \ decreases significantly of metal-poor stars. This can also be seen when comparing DR19 \cah \ values with those of APOGEE DR17 (Figure~\ref{mwmapogee}). The reliability range of \cah \ is shown in Table~\ref{abuerrorstab2}.

\subsection{Titanium}

The Ti abundance published in DR19 (\tih \,raw$\_$ti$\_$h) is derived from lines of neutral Ti. Our estimated precision is good, it varies between 0.073 and 0.149~dex in our three reference samples (Table~\ref{abuerrorstab} and Figure~\ref{abuerrors}). Ti is the least-precise $\alpha$-element published in DR19. While the zero-point corrections are excellent, the temperature dependence of \tih \ of the main-sequence stars in the SNSM sample is quite complex (see Ti panel of Figure~\ref{solarneigh}). This complexity is also visible in the MWM-GALAH common sample (Figure~\ref{mwmgalah}). For this reason, the \tih \ abundance of stars with \logg \ $>$ 3.8 is probably only reliable in a relatively narrow temperature range between 5500~K and about 4200$-$4300~K (Table~\ref{abuerrorstab2}). 

For the giants, Ti abundance can be reliably derived, but only in a narrow temperature region below 5500~K. Above this temperature, value the scatter is very large in the SNSM sample indicating the lack of real measurements. This large scatter at high temperatures (and also at low metallicities) is also visible when comparing the DR19 values with those of DR17 (Figure~\ref{mwmapogee}). Problems with giants also arise at temperatures around 4000~K and lower, but not for all stars. The Ti panel of Figure~\ref{abunmap} shows a group of stars that form a Ti-enriched population at solar metallicities, and the \tim \ values increase with \mh \ while the most metal-rich stars become even richer in Ti than the thick disk. This enrichment probably does not have any astrophysical origin, but some of them are associated with the group of C-rich cool stars defined in Sec.\ref{sect:quality_abund}. The exact reason of the complex correlation with \teff \ is still unknown; it is most likely related to the difficulty of fitting the often blended Ti lines at high metallicities in the top of the RGB. It is also possible that it is due to NLTE effects, which is why the overall quality of the Ti abundance in DR19 is only fair (Table~\ref{tab:abuerrorstab}). 

\subsection{Vanadium}

The abundance of vanadium (raw$\_$v$\_$h) is of poor quality, and it is one of the least-accurate and precise parameters published in DR19 because it is mainly derived from three very weak V lines. In fact, the best estimate of its precision is 0.265~dex in the solar neighborhood sample (Table~\ref{abuerrorstab} and Figure~\ref{abuerrors}). The distribution of \vh \ and \vm \ in the solar neighborhood (Figure~\ref{solarneigh}) and the Milky Way (Figure~\ref{abunmap}) shows multimodality at all temperatures, and the MWM-GALAH comparison displays significant temperature dependence. Since the V lines are very weak in the $H$ band, most of our measurements are unreliable (see the comparison with APOGEE DR17 V abundances in Figure~\ref{mwmapogee}), and careful assessment of the individual spectrum fittings is required before any science applications. \citet{2022ApJS..262...34H} carried out such analysis and determined \vm \ values using APOGEE spectra for several thousands of stars.

\subsection{Chromium}

The accuracy and precision of the chromium abundance (raw$\_$cr$\_$h) are excellent and good in DR19, respectively (Table~\ref{abuerrorstab2}). The zero-point corrections are moderate for both giants (0.032~dex) and dwarfs (0.069~dex), the estimated precision is between 0.077 and 0.178~dex, though this later value is based on the wide binary sample, which mostly consists of subgiants and dwarfs (Table~\ref{abuerrorstab} and Figure~\ref{abuerrors}). The poor precision and accuracy of \crh \ of main-sequence stars can be seen in the dwarf Cr panel of Figure~\ref{solarneigh}, in which a large scatter is visible along with a bimodality of Cr abundances below about 4200$-$4300~K. The noding effect seen in the MWM-APOGEE common sample (Figure~\ref{mwmapogee}) below 4500~K is also driven by dwarf stars. Thus, \crh \ in cool main-sequence stars may only be reliable above this temperature up to about 6000~K, where the scatter significantly increases. 

In giants, measuring Cr is much easier, so Cr abundances are reliable and have an estimated precision better than 0.1~dex up from 3700~K to about 6000~K, but above this temperature, the scatter of \crh \ increases significantly because the Cr lines become too weak. The expected chemical shape of the Milky Way is visible in Figure~\ref{abunmap}, with some slight multimodality seen in the metal-poor region. The MWM-GALAH comparison is fair with a 0.21~dex scatter, but this increases at low metallicities. The agreement between MWM DR19 and APOGEE DR17 values of \crh \ at low metallicities is also poor (Figure~\ref{mwmapogee}). Thus, \crh \ values of giant stars below about \mh \ $< -$1.0~dex may have larger errors than their solar-metallicity counterparts, which results in multiple stripes visible in the \crm \ vs. \mh \ distribution (Figure~\ref{abunmap}). Taking these factors into account, the abundance of Cr in DR19 is generally good.

\subsection{Manganese}

Even though the zero-point offsets applied for the Mn abundances (raw$\_$mn$\_$h) in DR19 are of fair quality and quite high at 0.134 and 0.144~dex for giants and dwarfs, respectively, its overall quality is still good. The precision is among the best of the elements, as our estimated uncertainties vary between only 0.035 and 0.077~dex (Table~\ref{abuerrorstab} and Figure~\ref{abuerrors}). The low scatter results in a narrow Mn distribution of the thin disk in the Milky Way (Figure~\ref{abunmap}). MWM DR19 and APOGEE DR17 manganese abundances agree very well, except perhaps at low metallicities (Figure~\ref{mwmapogee}).

Mn abundances are reliable in both giants and dwarfs, but in different temperature ranges (Table~\ref{abuerrorstab2}). In main-sequence stars, \mnh \ increases below 4000~K compared to the values at higher temperatures (which show a constant value) in the SNSM sample (Figure~\ref{solarneigh}), which is probably due to correlated errors coming from incorrect surface gravity determination (see Section 4.2 for more details). On the hot end, dwarf Mn abundances are reliable up to 6500$-$7000~K. In giants, the upper \teff \ limit is about 6000~K above which Mn abundances have large uncertainties due to weakening of the Mn lines. The MWM-GALAH sample shows a complex correlation with \teff \ and \mh, which appears to come from both optical and IR measurements, because we observe a temperature dependence of \mnh \ in the SNSM sample for the coolest dwarf stars below 4000~K (Figure~\ref{solarneigh2}).

\subsection{Cobalt}

Cobalt abundances (raw$\_$co$\_$h) are reliable in only a narrow parameter range in giants below 6000~K and above roughly 3700~K (Table~\ref{abuerrorstab2}). This is because \coh \ is determined from one absorption line in the $H$ band, which is fairly weak. The accuracy is generally good with a zero-point offset of $-$0.069~dex for giants. Dwarf cobalt abundances are unreliable indicated by their high scatter and multimodality observed in the SNSM sample (Figure~\ref{solarneigh}) which also makes the \coh \ vs. \mh \ distribution fairly noisy (as seen in Figure~\ref{abunmap}). The MWM-GALAH comparison also shows high scatter (Figure~\ref{mwmgalah}), along with the MWM-APOGEE common sample (Figure~\ref{mwmapogee}). The estimated uncertainty is highest in wide binaries (0.393~dex), and lowest in the SNSM sample, 0.144~dex; thus, the overall quality of \coh \ in DR19 is only fair.

\subsection{Nickel}

Nickel is among the most precisely measured elements in DR19 (raw$\_$ni$\_$h), our estimated overall precision is only between 0.027 and 0.038~dex (Table~\ref{abuerrorstab} and Figure~\ref{abuerrors}). The \nih \ vs. \mh \ exhibits a narrow distribution in the Milky Way even down to very low metallicites. The scatter starts to slightly increase below \mh \ $< -$2.0~dex (Figure~\ref{abunmap}), and the same low scatter can be seen in the SNSM sample too (Figure~\ref{solarneigh}). From this later sample, we can conclude that \nih \ is reliably measured in dwarfs in a large temperature range above 3500~K and below 7000~K. For giants, this range changes to 3700$-$6500~K; however, below about 4200$-$4300~K significant noding effect is visible when \nim \ is smaller than $-$0.1~dex (Figure~\ref{abuflags}). This noding of abundances around certain values is also visible when the DR19 Ni abundances are compared to those of DR17 (Figure~\ref{mwmapogee}). The MWM-GALAH common sample (Figure~\ref{mwmgalah}) does not show a significant offset, but only a slight temperature-dependent correlation, and the overall scatter (0.11~dex) is considered low. Thus, the \nih \ abundance in DR19 is of excellent quality.

\subsection{Copper}

Unfortunately, deriving abundances of copper (raw$\_$cu$\_$h) was not successful in DR19, resulting in high scatter of 0.207$-$0.357~dex in the solar neighborhood, open clusters, and wide binaries (Table~\ref{abuerrorstab} and Figure~\ref{abuerrors}), large zero-point offsets for both giants (0.323~dex) and dwarfs (0.214~dex). The obvious multimodalities are visible in the SNSM sample (Figure~\ref{solarneigh}) and the Milky Way (Figure~\ref{abunmap}), and a large offset exists between the MWM and GALAH \cuh \ values (Figure~\ref{mwmgalah}). It is possible to derive copper abundances from the H-band spectra; however, they require careful analysis of the two Cu lines that appear between 1.51 and 1.69 \mum \ \citep{2022ApJS..262...34H}

\subsection{Cerium}

The abundance of cerium (raw$\_$ce$\_$h) is determined from one line, which results in a fairly large scatter (0.225$-$0.231~dex, Table~\ref{abuerrorstab} and Figure~\ref{abuerrors}) in the solar neighborhood, open cluster stars and wide binaries. In fact, \ceh \ is mostly reliable (with high uncertainty) in giants only below 6000~K and above 4000~K (Figure~\ref{solarneigh}) when the metallicity is larger than about $-$1.5, and unreliable for all main-sequence stars in DR19. The \cem \ vs. \mh \ is fairly continuous; however, some noding effect can be seen for stars with \mh \ $>$ 0.0~dex at \cem \ around $-$0.2~dex, which probably does not have an astrophysical origin (Figure~\ref{abunmap}). This group of stars also shows up in the \xm \ vs. \teff \ plane below 4000~K showing multiple blobs and stripes for almost all stars in that temperature region (Figure~\ref{abuflags}). The differences between the DR19 and DR17 \ceh \ values also show these stripes at low temperatures along with a relatively large scatter at higher temperatures (Figure~\ref{mwmapogee}). The agreement with the optical GALAH \cem \ values shows a relatively large scatter of 0.32~dex. Overall, the quality of \ceh \ abundance in DR19 is fair.

\subsection{Neodymium}

The neodymium abundance (raw$\_$nd$\_$h) is derived from multiple fairly weak lines, which mostly appear in the spectra of stars with metallicity higher than about $-$1.5~dex; thus, its estimated uncertainty lies between 0.125 and 0.223~dex (Table~\ref{abuerrorstab} and Figure~\ref{abuerrors}). This scatter is visible in the Nd panel of Figure~\ref{solarneigh} and also accompanied by relatively large zero-point offsets (Table~\ref{zeropoint}) and multimodality of \ndm \ values of both main-sequence and giant stars with \teff \ $<$ 4200$-$4300~K (Figure~\ref{abuflags}). The multimodality also appears in the \ndh \ vs. \mh \ panel of Figure~\ref{abunmap}, though the thin disk is moderately defined. On the high end, Nd abundances can be measured up to about 6000~K for giants and 5500~K for dwarfs. The agreement with the optical GALAH measurements is also poor (Figure~\ref{mwmgalah}). Thus, Nd abundances should be used with care and are generally of fair quality in DR19.


\section{Conclusions}

ASPCAP is used for the primary published atmospheric parameters and abundances for nearly 900,000 stars in DR19. Our goal was to examine the accuracy and precision of these parameters, and our main results and conclusions can be summarized as follows:

\begin{enumerate}

\item \textit{Effective temperatures}: \teff \ values were compared to three different photometric temperatures scales. The median difference below for \logg \ $<$ 3.8 varies between $-$63.4 and $-$80.2 K, which is a slight discrepancy between the ASPCAP and photometric temperature, and for dwarfs, it is between $-$13.3 and 13.3 K, which is an excellent agreement. Based on the scatter of the differences, we found that the overall uncertainty of ASPCAP temperatures is around 50$-$70 K for giants, and 70$-$100 K for main-sequence stars.

\item \textit{Surface gravity}: The asteroseismic \logg \ values from the APOKASC and TESS catalogs served as the main comparison sample in DR19. We found systematic median offsets of 0.09 dex for RGB stars, 0.18 dex for RC stars, and 0.14 dex in the full TESS sample. While the accuracy is not perfect, the precision of ASPCAP surface gravities is excellent, varying between 0.07 and 0.09 in the APOKASC sample for RGB and RC stars, respectively, and somewhat worse in the TESS sample at 0.11~dex.

\item \textit{Metallicity}: Based on the common sample between MWM, GALAH, Gaia-ESO, and the Gaia Benchmark Stars, we found an overall excellent agreement between the various metallicity scales: the median differences are between 0.0 and 0.02~dex. The only discrepancy observed is in the MWM-GALAH common stars below 4500~K that also depends on metallicity. Based on the scatter observed in 12 open and 16 globular clusters, we estimate that the uncertainty of \mh \ in DR19 is better than 0.033~dex near solar metallicity and 0.092~dex below \mh \ $< -$1~dex.

\item \textit{Abundances}: Abundances of 21 elements are published in DR19. We examined their accuracy and precision using SNSM stars, open clusters, wide binaries, and the MWM-GALAH common sample. Abundances were zero-point corrected in DR19 using the solar neighborhood sample, and we provide temperature-based corrections using stars in open clusters as calibrators. We find that there are 10 elements $-$ metallicity, $\alpha$, C, N, O, Mg, Si, Ca, Fe and Ni $-$ that have excellent quality in DR19 with precision varying between 0.02 and 0.1~dex and low zero-point offsets. Al, S, K, Cr and Mn have good quality: these abundances have larger zero-point offsets and poorer precision and might show trends with temperature in the open cluster and/or SNSM sample. Na, Ti, Co, Ce, and Nd are considered to have poor quality because we were only able to measure them reliably in a narrow parameter region, and may exhibit noding around certain abundance values in the \xm \ vs. \teff, and/or \xm \ vs. \mh \ space. Measuring the abundances of P, V and Cu was unfortunately not successfull in DR19.

\end{enumerate}

\acknowledgements{
This project has been supported by the LP2021-9 Lend\"ulet grant of the Hungarian Academy of Sciences. P.J. acknowledges FONDECYT Regular grant No. 1231057. The research leading to these results has (partially) received funding from the KU\,Leuven Research Council (grant C16/18/005: PARADISE), from the Research Foundation Flanders (FWO) under grant agreement G089422N, as well as from the BELgian federal Science Policy Office (BELSPO) through PRODEX grant PLATO. P.F. and J.O. were supported by the National Science Foundation grant AST- 2206541. This project was supported by the NKFIH excellence grant TKP2021-NKTA-64. On behalf of the "New 3D Model Atmospheres for the SDSS Milky Way Mapper Survey" project, we are grateful for the possibility to use HUN-REN Cloud (see \citealt{heder2022}; https://science-cloud.hu/), which helped us achieve the results published in this paper. G.G. acknowledges support by Deutsche Forschungs-gemeinschaft (DFG, German Research Foundation) – project-IDs: eBer-22-59652 (GU 2240/1-1 "Galactic Archaeology with Convolutional Neural-Networks: Realising the potential of Gaia and 4MOST"). This project has received funding from the European Research Council (ERC) under the European Union’s Horizon 2020 research and innovation programme (grant agreement No. 949173). 

Funding for the Sloan Digital Sky Survey V has been provided by the Alfred P. Sloan Foundation, the Heising-Simons Foundation, the National Science Foundation, and the Participating Institutions. SDSS acknowledges support and resources from the Center for High-Performance Computing at the University of Utah. The SDSS web site is \url{www.sdss.org}.

SDSS is managed by the Astrophysical Research Consortium for the Participating Institutions of the SDSS Collaboration, including the Carnegie Institution for Science, Chilean National Time Allocation Committee (CNTAC) ratified researchers, the Gotham Participation Group, Harvard University, Heidelberg University, The Johns Hopkins University, L’Ecole polytechnique f\'{e}d\'{e}rale de Lausanne (EPFL), Leibniz-Institut f\"{u}r Astrophysik Potsdam (AIP), Max-Planck-Institut f\"{u}r Astronomie (MPIA Heidelberg), Max-Planck-Institut f\"{u}r Extraterrestrische Physik (MPE), Nanjing University, National Astronomical Observatories of China (NAOC), New Mexico State University, The Ohio State University, Pennsylvania State University, Smithsonian Astrophysical Observatory, Space Telescope Science Institute (STScI), the Stellar Astrophysics Participation Group, Universidad Nacional Aut\'{o}noma de M\'{e}xico, University of Arizona, University of Colorado Boulder, University of Illinois at Urbana-Champaign, University of Toronto, University of Utah, University of Virginia, Yale University, and Yunnan University.

This publication made use of NASA’s Astrophysics Data System.

}

\bibliographystyle{aa}
\bibliography{references}

\end{document}